\documentclass[notoc]{JHEP3}

\usepackage{cite}
\usepackage{epsfig}

\newcommand{\beq}{\begin{equation}}
\newcommand{\eeq}{\end{equation}}
\newcommand{\bea}{\begin{eqnarray}}
\newcommand{\eea}{\end{eqnarray}}
\newcommand{\nn}{\nonumber}

\def\eqn#1{Eq.~(\ref{#1})}
\def\eqns#1#2{Eqs.~(\ref{#1}) and~(\ref{#2})}

\def\fig#1{Fig.~{\ref{#1}}}
\def\sec#1{Section~{\ref{#1}}}
\def\app#1{Appendix~\ref{#1}}

\newcommand\fverb{\setbox\pippobox=\hbox\bgroup\verb}
\newcommand\fverbdo{\egroup\medskip\noindent%
                        \fbox{\unhbox\pippobox}\ }
\newcommand\fverbit{\egroup\item[\fbox{\unhbox\pippobox}]}
\newbox\pippobox

\def\spa#1.#2{\left\langle#1\,#2\right\rangle}
\def\spb#1.#2{\left[#1\,#2\right]}

\def\feynsl#1{
  \setbox0=\hbox{/} \setbox1=\hbox{$#1$}
  \dimen0=\wd0 \advance\dimen0 by -\wd1 \divide\dimen0 by 2
  \ifdim\wd0>\wd1 \raise.15ex\copy0\kern-\wd0\kern\dimen0\copy1\kern\dimen0
  \else \kern-\dimen0\raise.15ex\copy0\kern-\dimen0\kern-\wd1\copy1\fi}

\def\tr{\mathop{\rm tr}\nolimits}
\def\top{{\rm top}}
\def\mt{M_\top}
\def\eff{{\rm eff}}
\def\gev{{\rm GeV}}
\def\ord{{\cal O} }

\def\cL{{\cal L}}
\def\la{\langle}
\def\ra{\rangle}

\def\ib{{\bar\imath}}
\def\jb{{\bar\jmath}}
\def\qb{{\bar q}}
\def\Qb{{\overline Q}}

\newcommand\sss{\scriptscriptstyle}
\newcommand\as{\alpha_{\sss S}} 

\def\CF{C_{\sss F}}
\newcommand\ph{p_{\sss H}}
\newcommand\pt{p^{\sss\rm T}}
\newcommand\ptj{p^{\sss\rm T}_j}
\newcommand\ptmin{p^{\sss\rm T}_{min}}
\newcommand\mh{M_{\sss H}}

\newcommand\yrel{y_{\rm rel}}

\newskip\humongous \humongous=0pt plus 1000pt minus 100pt

\newif\ifdtup

\def    \br(#1,#2)          {\mbox{$\langle #1 \, #2 \rangle$}}
\def    \sq(#1,#2)          {\mbox{$\left[  #1 \, #2 \right]$}}

\title{Higgs Boson Production in Association with Three Jets}

\author{Vittorio Del Duca\\ 
Istituto Nazionale di Fisica Nucleare, Sez. di Torino\\
via P. Giuria, 1 - 10125 Torino, Italy\\
        E-mail: \email{delduca@to.infn.it}}

\author{Alberto Frizzo\thanks{Present Address: 
I.T.C.G. "A. Ceccato", via Vanzetti 14, 36016 Thiene (Vi) Italy}\\
Dipt. di Fisica Teorica, Universit\`a di Torino\\
via P. Giuria, 1 - 10125 Torino, Italy\\
        E-mail: \email{frizzo@to.infn.it}}

\author{Fabio Maltoni\thanks{Mail Address: Dipartimento di Fisica,
Terza Universit\`a di Roma, via della Vasca Navale, 84 - 00146
Rome, Italy.}\\
Centro Studi e Ricerche ``Enrico Fermi'' \\
via Panisperna, 89/A - 00184 Rome, Italy \\
E-mail: \email{maltoni@fis.uniroma3.it}}

\abstract{The scattering amplitudes for Higgs $+$~5 partons are computed,
with the Higgs boson produced via gluon fusion
in the large $\mt$ limit. A parton-level analysis
of Higgs $+$~3~jet production via gluon fusion and via weak-boson
fusion is presented, and the effectiveness of a central-jet veto is 
analysed.}

\preprint{{~DFTT 11/2004}}

\begin{document}

\section{Introduction}

The mechanism that governs the electroweak symmetry breaking is at
present the largest unknown in the Standard Model (SM) of elementary
particle physics. The canonical mechanism, the Higgs model, is a
keystone of the SM and its supersymmetric extensions. However, it is
based on the existence of a CP-even scalar particle, the Higgs boson,
which has not been detected yet and is the most wanted particle of the
Fermilab Tevatron and the CERN Large Hadron Collider (LHC) physics
programmes.  The direct search in the $e^+e^-\to ZH$ process at the
CERN LEP2 collider has posed a lower bound of 114.1~GeV on the SM Higgs mass, 
$\mh$~\cite{Barate:2000ts,Acciarri:2000hv,Abbiendi:2000ac,Abreu:2000fw,:2001xw}.
On the other hand, a SM Higgs boson with a mass of the order of 200~GeV
or less is favoured by analyses of the electroweak precision
data~\cite{Gambino:2003xc}.  Thus the
intermediate mass region ($\mh \lesssim$ 200 GeV) seems to be preferred.

\begin{figure}[h]
\hbox to\hsize{\hss
\includegraphics[width=.4\hsize]{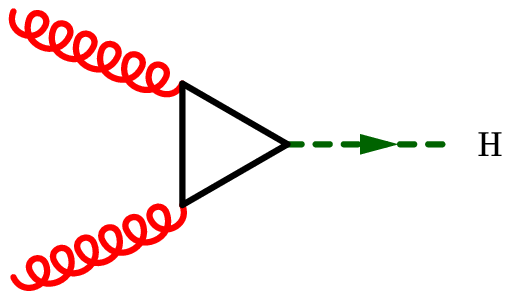}
\includegraphics[width=.4\hsize]{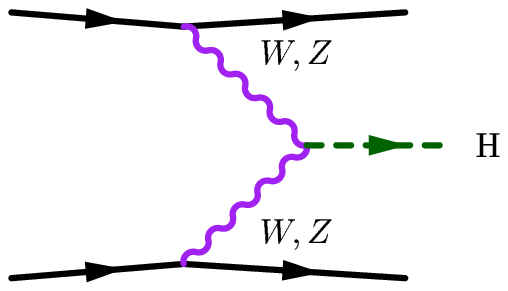} 
\hss}
\caption{Higgs production via gluon fusion (left) and via WBF (right).}
\label{fig:fig1}
\end{figure}

In $pp$-collisions, the dominant production mechanism over the entire
Higgs mass range accessible at LHC is via gluon fusion $g g\to
H$~\cite{Cavalli:2002vs}, where the coupling of the Higgs to the
gluons is mediated by a heavy quark loop, (left-hand panel of 
Fig.~\ref{fig:fig1}).  In
the SM, the leading contribution comes from the top quark, the
contributions from other quarks being at least smaller by a factor
$\ord(M_{\rm b}^2/\mt^2)$.  The second largest production mechanism is
via weak-boson fusion (WBF), $q q\to q q H$ (right-hand panel of
Fig.~\ref{fig:fig1})\footnote{In terms of significance, 
{\it i.e.} ratio of signal over root
of background $S/\sqrt{B}$, Higgs production via WBF may be larger
than the one via gluon fusion~\cite{Asai:2004ws}.}. In addition, it is a
key component of the program to measure the Higgs boson couplings at
LHC~\cite{Zeppenfeld:2000td}.

Higgs production via WBF occurs as the scattering
between two (anti)quarks with weak-boson ($W$ or $Z$) exchange in the
$t$-channel and with the Higgs boson radiated off the weak-boson
propagator.  Since the distribution
functions of the incoming valence quarks peak at values of the
momentum fractions $x\approx 0.1$ to 0.2, Higgs production via WBF
tends to produce naturally two highly energetic outgoing quarks. In
addition, the weak-boson mass provides a natural cutoff on the
weak-boson propagator, thus two jets with a transverse energy
typically of the order of a fraction of the weak-boson mass can be
easily produced. Therefore a large fraction of the events of the total
cross section for Higgs production via WBF occurs with two jets with a
large rapidity interval between them, typically one at forward and the
other at backward rapidities.
By considering Higgs production in association with two jets, and by
selecting events with a large dijet invariant mass and with a large
rapidity separation between the jets, the WBF production channel can
be made larger than the gluon fusion
one~\cite{DelDuca:2001eu,DelDuca:2001fn}.

In Refs.~\cite{DelDuca:2001eu,DelDuca:2001fn}, Higgs $+$~2~jet events have 
been analysed at leading order in $\as$. A broader dijet mass distribution, 
together with a tendency to have larger rapidity intervals between the two 
jets, for WBF than for gluon fusion was found. However, Higgs $+$~2~jet 
production via gluon fusion features a strong dependence on the 
renormalisation scale, because it is a calculation to $\ord(\as^4)$. 
The next-to-leading order (NLO) 
corrections to Higgs $+$~2~jet production via WBF are known and have been 
found to be typically modest, of the order of 5 to 10\%~\cite{Figy:2003nv}. 
Thus Higgs production via WBF is under a good theoretical control.
NLO corrections to Higgs $+$~2~jet production 
via gluon fusion are not known. Such a calculation would be quite formidable.
In fact, since the Higgs boson is produced via a
loop, a calculation of the production rate is quite involved,
even at leading order in $\as$: in Higgs $+$~2~jet production via gluon fusion
up to pentagon loops occur. Evaluating the NLO corrections would require
to calculate amplitudes with up to either hexagon loops or 
pentagon double loops.
On the other hand, it is known that the NLO corrections to the total
rate of Higgs production via gluon fusion are large, increasing the
production rate by up to 80\%~\cite{Graudenz:1992pv,Spira:1995rr}, and
so are the NLO corrections to Higgs $+$~1~jet production (computed in
the large $\mt$ 
limit)~\cite{deFlorian:1999zd,Ravindran:2002dc,Glosser:2002gm,mcfm}.
Thus, we are led to expect that also the unknown NLO corrections to
Higgs $+$~2~jet production be large. 

Even though a full NLO calculation of Higgs $+$~2~jet production via gluon 
fusion is not feasible with the present technology, it could be done
in the large top-quark mass limit, $\mh^2 \ll 4 M_\top^2$, whereby the
gluon-Higgs coupling via the top-quark loop is given by a low-energy theorem 
through an effective Lagrangian~\cite{Shifman:1979eb,Ellis:1976ap},
\beq
\cL_\eff = {1\over 4} A H G^a_{\mu \nu} G^{a~\mu \nu}\, ,\label{efflag}
\eeq
with $G^a_{\mu \nu}$ the field strength of the
gluon field and $H$ the Higgs-boson field (the effective coupling $A$
is given by $A = \as /(3 \pi v)$, where $v$ is the vacuum expectation
value parameter, $v^2=(G_F\sqrt{2})^{-1}=(246~\gev)^2$). For Higgs
production in association with jets, one must require in addition
that the jets and Higgs transverse energies are smaller than 
the top-quark mass~\cite{Baur:1989cm,DelDuca:2003ba}. 

In this work, we present the tree-level scattering amplitudes for the process
Higgs $\to 5$ partons, with the Higgs boson coupling to gluons through the
effective vertex of the large $\mt$ limit. The amplitudes have been
computed at fixed parton helicities. These amplitudes can be used to
compute Higgs $+$~3~jet production via gluon fusion, which we will analyse
in \sec{sec:higgstre}; they contribute to the radiative corrections of the
inclusive production of Higgs $+$~2~jets to NLO;
they can also be used to complete the calculation of the collinear 
splitting functions of a gluon into four partons (only the
splitting functions of a gluon into four gluons are 
known~\cite{DelDuca:2000ha}). The splitting functions of a parton into 
four partons are some of the building blocks of a recently proposed
method to compute the Altarelli-Parisi kernels to
next-to-next-to-leading order accuracy~\cite{Kosower:2003np}.

The paper is organised as follows: 
in the Introduction we have described the basic features of Higgs $+$~2~jet 
production via WBF and via gluon fusion, which motivate our choice of the 
final-state topology in Higgs $+$~3~jet production;
in \sec{sec:h5parton} we introduce
the scattering amplitudes for Higgs $+$~5 partons, whose explicit
expressions are collected in \app{sec:toplimhjjj};
in \sec{sec:higgstre} we present Higgs $+$~3~jet 
production via gluon fusion and via WBF;
in \sec{sec:concl} we present our conclusions.

\section{Amplitudes for Higgs $+$ 5 partons}
\label{sec:h5parton}

The scattering amplitudes for Higgs $+$~5 partons, with the Higgs boson 
coupling to gluons through the
effective vertex of the large $\mt$ limit, have been
computed at fixed parton helicities, and have been decomposed into
colour-stripped gauge-invariant sub-amplitudes,
whose explicit expression is reported in \app{sec:toplimhjjj}.
Here we show the colour decomposition of the amplitudes.

Since the Higgs boson is a colour singlet, the colour structure of
the QCD amplitudes for Higgs $+~n$ partons 
is the same as the one of the tree $n$-parton amplitudes~\cite{Mangano:1990by}.
We repeat it here for later convenience.

The colour decomposition of the tree $n$-gluon amplitudes 
is~\cite{Mangano:1990by,DelDuca:2000ha,DelDuca:2000rs}~\footnote{We use 
the standard normalisation of the fundamental representation matrices,
$\tr(T^a T^b) = \delta^{ab}/2$.
This is the origin of the factor $2^{n/2}$ which we make explicit
rather than carrying it over in the sub-amplitudes,
as in Ref.~\cite{DelDuca:2000ha}. The additional factor 2 is due to
the effective coupling of the Higgs to the gluons.
For $n= 4$, \eqn{GluonDecompNew} reproduces the 
normalisation of the sub-amplitudes of Ref.~\cite{Dawson:1992au}.} 
\begin{eqnarray}
\lefteqn{ {\cal M}_n(1,\ldots,n) }\nonumber\\ &=& 2^{(n-2)/2} g^{n-2} 
\sum_{\sigma\in S_n/{\Bbb Z}_n}
   \tr (T^{a_{\sigma_1}}\cdots T^{a_{\sigma_n}})
    m_n(\sigma_1,\ldots,\sigma_n) \label{GluonDecompNew}\\
&=& 2^{(n-2)/2} \frac{(ig)^{n-2}}{2} \sum_{\sigma\in S_{n-2}}
      f^{a_1 a_{\sigma_2} x_1} f^{x_1 a_{\sigma_3} x_2} \cdots f^{x_{n-3} 
a_{\sigma_{(n-1)}} a_n}
m_n(1,\sigma_2,\ldots,\sigma_{n-1},n)\, ,\nn
\end{eqnarray}
where $S_n/{\Bbb Z}_n$ are the non-cyclic permutations of the $n$ gluons,
and $A$ is given below \eqn{efflag}.
The dependence on the particle helicities and momenta in the sub-amplitude,
and on the gluon colours in the trace, is implicit
in labelling each leg with the index $i$.
Helicities and momenta are defined as if all particles were outgoing.
The gauge invariant sub-amplitudes $m$
satisfy the relations~\cite{Berends:1988me},
\begin{eqnarray}
&& m(1,2,\ldots,n-1,n) = m(n,1,2,\ldots,n-1) 
   \hspace{3.5cm}  {\rm cyclicity} \nonumber \\
&& m(1,2,\ldots,n) = (-1)^n m(n,\ldots,2,1)  
   \hspace{4.8cm} {\rm reflection}  \label{relations}\\
&& m(1,2,3,\ldots,n) + m(2,1,\ldots,n)+\ldots + m(2,3,\ldots,1,n) = 0
   \hspace{.5cm} {\rm dual \;Ward \;identity}\, . \nonumber 
\end{eqnarray}

The colour decomposition of the tree amplitudes for $(n-2)$ gluons 
and a $q\bar q$ pair is
\begin{equation}
 {\cal M}_n(q,\bar{q}; 3,\ldots,n)
  = 2^{(n-4)/2} g^{n-2} \sum_{\sigma\in S_{(n-2)}}
   (T^{a_{\sigma_3}}\ldots T^{a_{\sigma_n}})_{i}^{~\jb}\
    m_n(1_q,2_{\bar{q}}; \sigma_3,\ldots,\sigma_n)\, ,
\label{TwoQuarkGluonDecomp}
\end{equation}
where $S_{(n-2)}$ is the permutation group of the $(n-2)$ gluons.
Reflection symmetry is like in \eqn{relations}, for gluons and/or quarks
alike. 

The colour decomposition of the tree amplitudes for two $q\bar q$ pairs
and $(n-4)$ gluons is
\begin{eqnarray}
\lefteqn{ {\cal M}_n(1_q,2_\qb,3_Q,4_{\overline{Q}};5,\ldots,n)
= 2^{(n-6)/2} g^{n-2} \sum_{k=0}^{n-4} \sum_{\sigma\in S_{k}}\,
\sum_{\rho\in S_{l}} }
\label{FourQuarkGluonDecomp}\\
&\times
& \Biggl[ (T^{a_{\sigma_1}} \dots T^{a_{\sigma_k}})_{i_3}^{\;\ib_2}\
  (T^{a_{\rho_1}} \dots T^{a_{\rho_l}})_{i_1}^{\;\ib_4}\
    m_n(1_q,2_\qb,3_Q,4_{\overline{Q}};\sigma_1,\dots,
  \sigma_k;\rho_1,\dots,\rho_l) \nonumber\\
&+&
  {1\over N_c}\, (T^{a_{\sigma_1}} \dots 
T^{a_{\sigma_k}})_{i_1}^{\;\ib_2}\
  (T^{a_{\rho_1}} \dots T^{a_{\rho_l}})_{i_3}^{\;\ib_4}\
    {\tilde m}_n(1_q,2_\qb,3_Q,4_{\overline{Q}};\sigma_1,\dots,
  \sigma_k ; \rho_1,\dots,  \rho_l) \Biggr]\, ,\nonumber
\end{eqnarray}
with $k+l=n-4$. The sums are over the partitions of $(n-4)$
gluons between the two quark lines, and over the permutations of the
gluons within each partition. For $k=0$ or $l=0$, the colour strings
reduce to Kronecker delta's. 

Several checks on the calculation were performed.  At the level
of colour ordered amplitudes we verified the gauge invariance, the generic
$1\to 3$ collinear limits, the $g \to gggg$ collinear 
limit~\cite{DelDuca:2000ha} and the $m_H, p_H \to 0$ limit where the
amplitudes reduce to the pure QCD five-parton amplitudes. At the level of
squared amplitudes summed over colour and helicity we
compared numerically with Alpha~\cite{Caravaglios:1995cd,Mangano:2002ea} 
and MadGraph~\cite{Stelzer:1994ta,Maltoni:2002qb} and
found agreement at the machine level precision.

\section{Higgs $+$ 3 Jet Production}
\label{sec:higgstre}

A distinguishing feature of WBF is that to leading order no colour 
is exchanged in the $t$-channel~\cite{Dokshitzer:1987nc,Bjorken:1993er}.
To $\ord(\as)$, gluon radiation occurs only as bremsstrahlung off the quark
legs (since no colour is exchanged in the $t$-channel in the Born process,
no gluon exchange is possible to $\ord(\as)$, except for
a tiny contribution due to equal-flavour quark scattering with 
$t\leftrightarrow u$ channel exchange). 

The different gluon radiation pattern expected for Higgs production via
WBF versus its major backgrounds, namely $t\bar{t}$ production and
QCD $WW + 2$~jet production, 
is at the core of the central-jet veto proposal, both for 
heavy~\cite{Barger:1994zq} and light~\cite{Kauer:2000hi} Higgs searches.
A veto of any additional jet activity in the central rapidity region
is expected to suppress more the backgrounds than the signal,
because the QCD backgrounds are characterised by quark or gluon exchange
in the $t$-channel. The exchanged partons, being coloured, are expected 
to radiate off more gluons.

\EPSFIGURE[t]{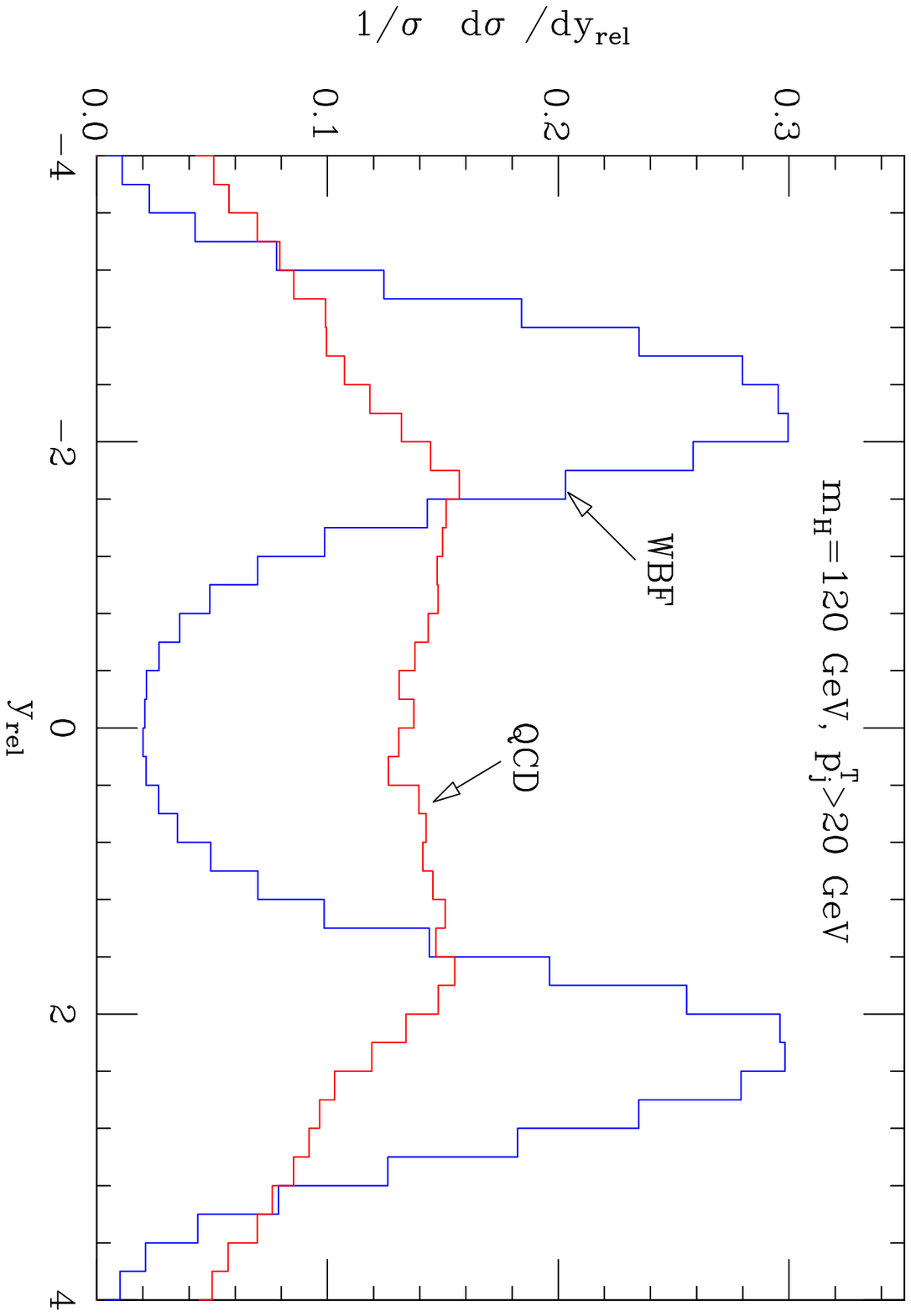,angle=90,width=.8\textwidth,clip=}
{The distribution in the rapidity of the third jet, 
measured with respect to the rapidity average of the tagging jets,
$y_{rel} = y_{j3} - (y_{j1}+y_{j2})/2$.
\label{fig:yrel} }

For the analysis of the Higgs coupling to gauge bosons,
Higgs $+$~2~jet production via gluon fusion may also be treated as a 
background.
When the two jets are separated by a large rapidity interval, the
scattering process is dominated by gluon exchange in the $t$-channel.
Therefore, like for the QCD backgrounds, the bremsstrahlung radiation 
is expected to occur everywhere in rapidity.\footnote{An analogous
difference in the gluon radiation pattern is expected in $Z +2$~jet
production via WBF fusion versus QCD production of 
$Z +2$~jets~\cite{Rainwater:1996ud}.}
In order to analyse that,
we consider Higgs $+$~3~jet production via WBF and via gluon fusion,
at LHC.\footnote{WBF results have been obtained with MadEvent~\cite{Maltoni:2002qb}.} 
On the three jets, we require the set of cuts
\beq \label{eq:cuts_min}
\ptj>20\;{\rm GeV}, \qquad |y_j|<5,\qquad R_{jj}>0.6,
\eeq
where $\ptj$ is the transverse momentum of a final state parton,
$y_j$ is its rapidity
and $R_{jj}$ is the separation of the two partons in 
rapidity versus azimuthal angle plane
\beq 
R_{jj} = \sqrt{\Delta y_{jj}^2 + \phi_{jj}^2}\;.
\eeq
In addition, like in Ref.~\cite{Figy:2003nv}, we select the two jets
with the highest transverse energy as the \emph{tagging jets} of our
events, and require them to pass the cuts
\beq \label{eq:cut_gap}
|y_{j1}-y_{j2}|>4.2, \qquad y_{j1}\cdot y_{j2}<0, \qquad
m_{jj}>600\;{\rm GeV},
\eeq
i.e.\ the two tagging jets must be well separated, must be in 
opposite detector hemispheres and must possess a large dijet 
invariant mass.
In \fig{fig:yrel}, we plot the distribution in the rapidity of the third jet, 
measured with respect to the rapidity average of the tagging jets,
$\yrel = y_{j3} - (y_{j1}+y_{j2})/2$, for gluon fusion and for WBF. 
Here and in \fig{fig:ptmin} the Higgs mass was set to $\mh = 120$~GeV, 
the CTEQ5L set of parton distribution functions~\cite{Lai:1994bb} 
was used along with the corresponding one-loop running of $\as$. 
The factorisation scales were set equal to the geometric average of the $\pt$
of the jets. About the choice of  the renormalisation scale, 
note that Higgs $+$~3~jet production via gluon fusion is an $\ord(\as^5)$
process, and a leading order process. Thus, its dependence on the
renormalisation scale is very large. Since Higgs $+$~3~jet production with
two tagging jets that pass the cuts (\ref{eq:cut_gap}) is dominated by
configurations with gluon exchange in the $t$ 
channel, a set of ``natural'' scale choices is given by the replacement 
$\as^5\to \as^2(\mh) \as(\pt_1) \as(\pt_2) \as(\pt_3)$, in analogy
with Higgs $+$~2~jet production with
two forward jets~\cite{DelDuca:2001fn,DelDuca:2003ba}.
However, in order to reduce the dependence on the scale choices, 
in \fig{fig:yrel} we normalise the distribution of 
the third jet to the Higgs $+$~3~jet production rate.
The different dynamical generation of the third jet is 
apparent: in WBF its production in the central rapidity region 
is naturally suppressed.

\EPSFIGURE[t]{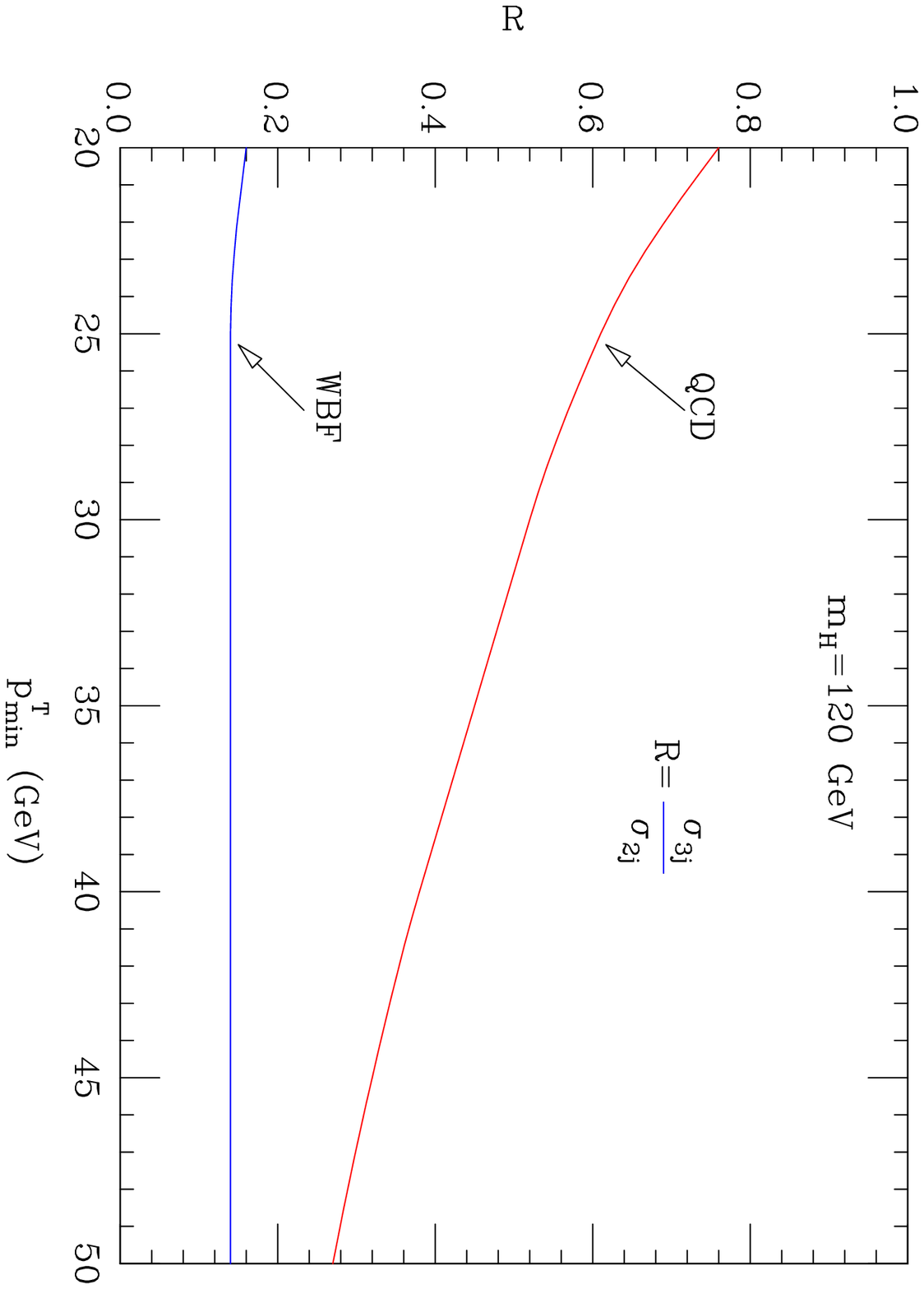,angle=90,width=.8\textwidth,clip=}
{The ratio of the Higgs $+$~3~jet production rate to Higgs $+$~2~jet, 
as a function of $\ptmin$.
\label{fig:ptmin} }

The central-jet veto discards all events with three or more jets
if any jet other than the tagging jets is produced above a given cut
$\ptmin$. Our jet selection implies then that the third largest jet
in $\pt$ be produced above $\ptmin$. 
For the central-jet veto proposal it is of interest to analyse
the ratio $R$ of the Higgs $+$~3~jet production rate to Higgs $+$~2~jet: 
the larger is $R$, the
more effective the central-jet veto is expected to be. In \fig{fig:ptmin},
we analyse $R$ as a function of $\ptmin$ for gluon fusion and for WBF.
Both Higgs $+$~3~jet and Higgs $+$~2~jet production were
evaluated to leading order in $\as$\footnote{For consistency with
Higgs $+$~3~jet, also the tagging jets of Higgs $+$~2~jet production have
been taken above the cut $\ptmin$.}.
We note that the gluon fusion process is monotonically dependent on $\ptmin$
while WBF is basically independent of it. Thus the effectiveness of the 
central-jet veto grows as $\ptmin$ decreases.

\section{Conclusions}
\label{sec:concl}

In this work we have computed the scattering amplitudes for Higgs $+$~5 
partons, with the Higgs boson produced via gluon fusion
in the large $\mt$ limit. Using those amplitudes, we have performed a 
parton-level analysis of Higgs $+$~3~jet production, via gluon fusion
and via WBF.
Since our ultimate goal is to collect information useful to study
the Higgs coupling to gauge bosons,
Higgs production via gluon fusion is treated as a background
that must be distinguished from WBF and contained as much as possible.
To that effect, in the literature forward jet tagging for Higgs $+$~2~jet 
production has been proposed, 
augmented by a central-jet veto for the production of a Higgs in association
with three or more jets. We have analysed the jet veto proposal by comparing
Higgs $+$~3~jet production via gluon fusion and via WBF. We have found that
the more severe the cut on the third jet, the more effective the jet veto.

\subsection*{Acknowledgments} 

VDD and FM thank Dieter Zeppenfeld for useful discussions and for his insight, 
Mauro Moretti for a numerical comparison on the amplitudes for Higgs $+$~5 
partons, and the Kavli Institute for Theoretical Physics
of the University of California at Santa Barbara for the kind hospitality
during the last stage of this work.  
This research was supported in part by the National Science
Foundation under Grant No. PHY99-07949.

\appendix

\section{Helicity amplitudes}
\label{sec:tophel}

We use helicity amplitudes,
defined in terms of massless Dirac spinors $\psi_{\pm}(p)$ of fixed helicity,
\begin{equation}
\psi_{\pm}(p) = {1\pm \gamma_5\over 2} \psi(p) \equiv 
|p^\pm\rangle\, , \qquad \overline{\psi_{\pm}(p)} \equiv \langle p^\pm|
\, ,\label{spi}
\end{equation}
spinor products,
\begin{equation}
\langle p k\rangle \equiv \langle p^- | k^+ \rangle\, , \qquad 
\left[pk\right] \equiv \langle p^+ | k^- \rangle\, , \qquad {\rm with}\;\;
\langle p k\rangle^* = {\rm sign}(p^0 k^0) \left[ k p\right]\, ,
\end{equation}
currents,
\bea
\langle i^-| p_{k_1} \dots p_{k_{2n+1}}| j^-\rangle &\equiv&
\langle i^-| \slash  \!\!\! p_{k_1} \dots \slash  \!\!\! p_{k_{2n+1}}  |j^-\rangle \nn\\
\langle i^-| p_{k_1} \dots p_{k_{2n}}| j^+\rangle &\equiv&
\langle i^-| \slash  \!\!\! p_{k_1} \dots \slash  \!\!\! p_{k_{2n}}  |j^+\rangle
\eea
Using the decomposition of the fermion propagator
\beq
\slash  \!\!\! p = |p^+\rangle \langle p^+| + |p^-\rangle \langle p^-| \, 
.\label{fermdecomp}
\eeq
the simplest current can be written as
\beq
\langle i^-| p_k | j^-\rangle = \langle i k \rangle \left[k j\right]
\eeq
and every other current can be written in terms of products of spinor products
by applying \eqn{fermdecomp} repeatedly on it.

In order to set up the 
sign and symmetry conventions of our amplitudes, we collect the known 
sub-amplitudes for Higgs $+$ three and four partons.

\subsection{Sub-amplitudes for Higgs + three partons}
\label{sec:toplimhj}

The colour decomposition of the tree amplitudes for Higgs plus three gluons 
is given in \eqn{GluonDecompNew}, with $n$ = 3. The independent
sub-amplitudes are~\cite{Kauffman:1997ix}
\begin{eqnarray}
&&m_3(1^+,2^+,3^+)= i A {\mh^4\over \la 1 2\ra \la 2 3\ra
   \la 3 1\ra} \, ,\nn \\  
&&m_3(1^-,2^+,3^+)= i A {[23]^3 \over [1 2] [13]}\, .
\label{mpp}
\end{eqnarray}
All of the other sub-amplitudes can be obtained by relabelling and
by use of reflection symmetry, \eqn{relations}, and parity inversion.
Parity inversion flips the helicities of all particles,
and it is accomplished by the substitution
$\spa{i}.j \leftrightarrow \spb{j}.i$.

The colour decomposition of the tree amplitudes for Higgs, a gluon
and a $q\bar q$ pair is given in \eqn{TwoQuarkGluonDecomp} for $n$ = 3.
There is only one independent sub-amplitude~\cite{Kauffman:1997ix}
\beq
m_3(1_q^-, 2_{\bar{q}}^+; 3^+) = 
i A  { [23]^2 \over [12] }\, .\label{qmpp}
\eeq
All other sub-amplitudes can be obtained by use of parity inversion and charge 
conjugation\footnote{In performing parity inversion, there is
a factor of $-1$ for each pair of quarks participating in the amplitude.}.
Charge conjugation swaps quarks and antiquarks without inverting helicities.

\subsection{Sub-amplitudes for Higgs + four partons}
\label{sec:toplimhjj}

The colour decomposition of the amplitudes for Higgs plus four gluons 
is given in \eqn{GluonDecompNew} for $n$ = 4. The independent
sub-amplitudes are~\cite{Dawson:1992au,Kauffman:1997ix}
\begin{eqnarray}
&&m_4(1^+,2^+,3^+,4^+)= i A {\mh^4\over \la 1 2\ra \la 2 3\ra
   \la 3 4\ra \la 4 1\ra} \, ,\nn\\ 
&&m_4(1^-,2^+,3^+,4^+) = i A
\left( {\la 1^-|\ph |3^-\ra^2 [24]^2 \over t_{124} s_{12} s_{14}}     
+ {\la 1^-|\ph | 4^- \ra^2 [2 3]^2 \over t_{123} s_{12} s_{23}} 
+ {\la 1^-|\ph | 2^- \ra^2 [3 4]^2 \over t_{134} s_{14} s_{34}} \right) \nonumber\\
&&\qquad   -i A {[2 4] \over 
[ 1 2 ] \la 2 3\ra \la 3 4 \ra   [ 4 1]} 
\biggl\{ s_{23} {\la 1^- |\ph | 2^- \ra \over \la 4 1\ra } + s_{34} 
{\la 1^- |\ph | 4^- \ra \over \la 1 2\ra }
-[2 4 ] t_{234}\biggr\} \, ,
\label{mppp}\\ 
&&m_4(1^-,2^-,3^+,4^+)= iA \left( {\la 1 2\ra^4\over \la 12 \ra 
\la 23 \ra \la 34\ra \la 41\ra}
 + {[34]^4 \over [1 2] [23] [34] [41]} \right) \, ,
\nn
\end{eqnarray}
with the Higgs momentum in a current defined through momentum conservation
$\ph = - \sum_{i=1}^n p_i$.
All of the other sub-amplitudes can be obtained by relabelling and
by use of reflection symmetry, \eqn{relations}, and parity inversion.

The colour decomposition of the amplitudes for Higgs, two gluons
and a $q\bar q$ pair is given in \eqn{TwoQuarkGluonDecomp}, with $n$ = 4.
The independent sub-amplitudes are~\cite{Kauffman:1997ix}\footnote{For
the colour ordering on the fermion line we choose the convention of 
Ref.~\cite{Mangano:1990by},
which is the opposite of the one used in Ref.~\cite{Kauffman:1997ix}.},
\bea
m_4(1_q^+, 2_{\bar{q}}^-; 3^+, 4^+) &=& i A \left[
{ \la 2^-| \ph| 4^- \ra^2 \over t_{123} }
  { [13] \over \la 23 \ra } \left({1\over s_{12}} + {1\over s_{13}} \right)
  - { \la 2^-| \ph| 3^- \ra^2 \over t_{124} s_{12}} { [14] \over \la 24 \ra }
  - { \la 2^-| \ph| 1^- \ra^2 \over [12] \la 23 \ra \la 24 \ra \la 34 \ra }
\right] \nn \\
m_4(1_q^+, 2_{\bar{q}}^-; 3^-, 4^+) &=& -i A \left(
  {\la 23\ra^3 \over \la 12\ra \la 24\ra \la 34\ra } 
  - { [14]^3 \over [12][13][34] } \right) \, , \label{qpmpm} \\
m_4(1_q^+, 2_{\bar{q}}^-; 3^+, 4^-) &=& -i A \left( -
{ [13]^2 [23] \over [12][24][34] }
  + {\la 14\ra \la 24\ra^2 \over \la 12\ra \la 13\ra \la 34\ra } \right)
\nn
\eea
All of the other sub-amplitudes can be obtained by relabelling and
by use of parity inversion, reflection symmetry and charge 
conjugation.

The colour decomposition of the amplitudes for Higgs and two $q\bar q$ 
pairs is given in \eqn{FourQuarkGluonDecomp}, with $n$ = 4, which can be
re-cast as,
\beq
{\cal M}_4(1_q, 2_{\bar{q}}; 3_Q, 4_{\Qb}) = g^2 {1\over 2}
 \left( \delta_{i_3}^{\ib_2} \delta_{i_1}^{\ib_4}
- {1\over N_c}\delta_{i_1}^{\ib_2}\delta_{i_3}^{\ib_4}\right)
    m_4(1_q, 2_{\bar{q}}; 3_Q, 4_{\Qb})\, .\label{FourQuarkDecomp}
\eeq
There is one independent sub-amplitude~\cite{Kauffman:1997ix},
\beq
m_4(1_q^+, 2_{\bar{q}}^-; 3_Q^+, 4_{\Qb}^-) = -i A \left(
{\la 24\ra^2 \over \la 12\ra \la 34\ra} + { [13]^2 \over [12][34] } \right) \, 
.\label{hqqqqt}
\eeq
All the other sub-amplitudes, except for
$m_4(1_q^+, 2_{\bar{q}}^-; 3_Q^-, 4_{\Qb}^+)$
can be obtained by relabelling and
by use of parity inversion, reflection symmetry and charge 
conjugation. $m_4(1_q^+, 2_{\bar{q}}^-; 3_Q^-, 4_{\Qb}^+)$
is obtained from \eqn{hqqqqt} by inverting the parity on the current
$\la 3^-| \gamma^\mu |4^-\ra$. That amounts to swap the labels 3 and 4
in \eqn{hqqqqt}.

\section{Sub-amplitudes for Higgs $+$ five partons}
\label{sec:toplimhjjj}

In Appendices~\ref{sec:appba}, \ref{sec:appbb} and \ref{sec:appbc} we collect 
the colour-stripped sub-amplitudes for Higgs $+$ five partons. The squared 
amplitude for Higgs $+$ five gluons, summed over colours, is worked out
in \sec{sec:appc}.

\subsection{Sub-amplitudes for Higgs $+$ five gluons}
\label{sec:appba}

The colour decomposition of the amplitudes for Higgs plus five gluons 
is given in \eqn{GluonDecompNew} for $n$ = 5. 
There are three independent sub-amplitudes, we may take to be the ones
with zero, one and two gluons of negative helicity. They are 
\begin{eqnarray}
m_5(1^+,2^+,3^+,4^+,5^+)&=&i  A  \frac{M_{\sss H}^4}
{\br(1,2) \br(2,3) \br(3,4) \br(4,5) \br(5,1) } \, , 
\label{sec:Appppp}
\end{eqnarray}

\begin{eqnarray}
\lefteqn{m_5(1^-,2^+,3^+,4^+,5^+)=i  A  \left\{
\frac{\langle 1^-|(p_2+p_3)p_{\sss H}|1^+ \rangle ^2 }{\br(2,3) \br(4,5) }
\left[ \frac{\sq(2,5) }
{s_{12}s_{15}\br(1,3) \br(1,4)  }
\right. \right. } \nonumber \\
&& \left.  +\frac{\sq(3,2) }{s_{12}\br(1,4) \br(1,5)  t_{123}}-
\frac{\sq(4,5) }{s_{15} \br(1,2) \br(1,3) 
 t_{145}} \right] \nonumber \\
&&+\frac{ \langle 1^-|(p_2+p_5)p_{\sss H}|1^+ \rangle ^2 \sq(2,5) ^2}
{s_{12}s_{15}\br(1,3) \br(1,4) \br(3,4) t_{125}} \nonumber \\
&&-\frac{\langle 1^-|p_{\sss H}|2^- \rangle ^2}
{s_{15} \br(1,3) \br(3,4) \br(4,5) \sq(1,2) }\left[\sq(2,5) +
\frac{\br(3,4) \sq(1,2) \sq(4,5) \langle 1^-|p_4+p_5|3^- \rangle }
{t_{145}s_{2{\sss H}}}\right] \nonumber  \\
&& -\frac{\sq(2,5) t_{234} \langle 1^-|p_{\sss H}|5^- \rangle ^2}
{s_{12} \br(1,4) \br (3,4) \br(2,3) \sq(1,5) s_{5{\sss H}}}
+\frac{ \sq(2,3)  \langle 1^-|p_{\sss H}|5^- \rangle ^2 \langle 1^-|p_2+p_3|4^-
\rangle }{s_{12} \br(1,4) \br(2,3) t_{123} s_{5{\sss H}}} \nonumber \\
&&+ \frac{\langle 1^-|p_{\sss H}|5^- \rangle ^3}{\br(1,2) \br(1,4) \br(2,3)
\br(3,4) \sq(1,5) s_{5{\sss H}}}
 \nonumber \\
&&-\frac{\langle 1^-|p_{\sss H}|4^- \rangle ^2 \langle 1^-|p_2+p_3|5^- \rangle }
{s_{12}\br(2,3) s_{4{\sss H}}}
\left[
\frac{\sq(5,2) }{s_{15}\br(1,3) }+
\frac{\sq(2,3) }{ \br(1,5)  t_{123}} \right]
\nonumber \\
&&-\frac{\sq(2,5) ^2 \langle 1^-|p_2+p_5|3^- \rangle \langle 1^-|p_{\sss H}|4^-
\rangle ^2}
{s_{12}s_{15}\br(1,3) t_{125}s_{4{\sss H}}} \nonumber \\
&&+\frac{\langle 1^-|p_{\sss H}|3^- \rangle ^2}{\br(1,2) s_{15} s_{3{\sss H}}}\left[
\frac{\langle 1^-|p_4+p_5|2^- \rangle }{\br(4,5) }\left( \frac{\sq(2,5) }{\sq(1,2)
\br(1,4)  }+\frac{\sq(4,5) }{t_{145}}\right)+
\frac{\sq(2,5) ^2 \langle 1^-|p_2+p_5|4^- \rangle }{\sq(2,1) \br(1,4)
t_{125}} \right] \nonumber \\
&&\left. -\frac{\langle 1^-|p_{\sss H}|2^- \rangle ^2 \langle 1^-|p_3+p_4|5^- \rangle }
{s_{15}\br(1,3) \br(3,4) \br(4,5) s_{2{\sss H}}} \right\} \, ,
\label{sec:Ampppp}
\end{eqnarray}

\begin{eqnarray}
\lefteqn{m_5(1^-,2^-,3^+,4^+,5^+)=i  A  \left\{
 \frac{-\br(1,2) ^2}{\br(1,3) \br(3,4) \br(4,5) \sq(1,2) s_{15}} [
\sq(1,2) \langle 1^-|p_3+p_4|5^- \rangle -\sq(2,5) t_{345} ] \right. }\nonumber \\
&&+\frac{1}{4 \br(1,3) \br(1,4) \br(4,5) \sq(1,2) \sq(2,5) s_{15}s_{23}t_{123}}
\left\{ 2\sq(3,5) \left[   2\br(1,2) ^2 \sq(2,5) \sq(3,5) (-s_{23}\br(1,5) \br(3,5) \sq(2,5) 
\right. \right. \nonumber \\
&&+\br(1,3) \sq(1,2) \sq(3,4) (\br(1,5) \br(3,4) +\br(1,4) \br(3,5) )) -
2 \br(1,3) \br(2,3) \sq(2,5) (s_{13}+s_{23}) \langle 1^-|p_4+p_5|3^- \rangle  ^2 \nonumber \\
&&+\br(1,2) \left( -2 \br(1,5) ^2\br(2,3) \sq(2,5) ^2 \sq(3,5) s_{23}-
2\br(1,3) ^2 \sq(2,5) \sq(3,4) \sq(3,5) \langle 5^-|p_2p_1-p_1p_2|4^+ \rangle 
\right. \nonumber \\
&&\left. \left. +
\br(1,3) \br(2,3) \left( 2\br(1,5) \sq(2,5) \sq(3,5) \langle 3^+|-p_2p_5-p_5p_1|2^- \rangle
+\br(1,4) ^2 \sq(1,2) \sq(3,4) (\sq(2,5) \sq(3,4) +\sq(2,4) \sq(3,5) )
\right. \right. \right. \nonumber \\
&&\left. \left. \left. +
\br(1,4) \br(1,5) \sq(1,2) \sq(3,5) (7\sq(2,5) \sq(3,4) +\sq(2,4) \sq(3,5) )\right) \right) \right]
 \nonumber \\
&&+\br(1,2) \sq(4,5) \left[ \br(1,2) \sq(2,5) s_{12} (\br(1,4) \sq(3,4) \langle 4^-|2p_1+p_5|2^- \rangle
+\br(1,5) \sq(3,5)  \langle 4^-|2p_1+2p_3+p_5|2^- \rangle ) \right. \nonumber \\
&&+\langle 1^-|p_4+p_5|3^- \rangle \left( -4 s_{23}^2 \br(1,4) \sq(2,5) +
\br(1,3) ^2 \sq(1,2) \sq(3,5)  \langle 4^-|-2p_1-p_5|3^- \rangle \right. \nonumber \\
&&\left. -2 s_{23} \br(1,3) (-2 \br(1,4) \sq(1,5) \sq(2,3) +\br(4,5)
\sq(2,5) \sq(3,5) ) 
\right) \nonumber \\
&&+2\br(1,2) \sq(2,3) (\br(1,3) \sq(1,2) (\br(1,4) ^2 \sq(1,5)
\sq(3,4) +\br(1,5) \sq(3,5) \langle 4^-|-p_1-p_2|5^- \rangle ) \nonumber
\\
&& \left. +\br(2,3) \sq(2,5) (\br(1,4) \sq(3,4) \langle 4^-|-3p_1-p_5|2^- \rangle 
+\br(1,5) \sq(3,5) \langle 4^-|-2p_1-p_5|2^- \rangle ) ) \right]
\nonumber \\
&&-\br(1,2) ^2 \br(1,4) \sq(2,3) \sq(4,5) ^2 (2 s_{23} \langle
4^-|p_1+p_5|2^- \rangle
+s_{12} \langle 4^-|2p_1+3p_5|2^- \rangle \nonumber \\
&&\left. +\br(1,3) \sq(1,2)  \langle 4^-|-2p_1-p_5|3^- \rangle )
\right\} \nonumber \\
&&+\frac{\sq(4,5) }{\br(1,3) \br(4,5) \sq(2,5) s_{15} s_{23}t_{145}}
\left[
-\br(1,3) \br(2,3) \sq(3,5)  \langle 1^-|p_4+p_5|3^- \rangle ^2
\right. \nonumber \\
&&+\br(1,2) \sq(3,5) (-\br(1,3) \br(1,4) \sq(3,4) s_{23}-\br(1,5) ^2
\br(2,3) \sq(2,5) \sq(3,5) -\br(1,5) \br(1,3) \sq(3,5)
(s_{23}+s_{25})) \nonumber \\
&&\left. -\br(1,2) ^2 \sq(2,5) (-s_{23}\br(1,4) \sq(3,4) -\br(1,5) \sq(3,5)
(s_{23}+s_{35}))-
\br(1,2) ^3 \sq(2,3) \sq(2,5) t_{145} \right] \nonumber \\
&&-\frac{\br(1,2)  \langle 1^-|p_3+p_4|5^- \rangle }{\br(1,3) \br(1,4)
\br(3,4) \sq(1,2) s_{15}t_{125}} [
-(s_{12}+s_{15})\langle 1^-|p_3+p_4|5^- \rangle +
\langle 5^+|p_2(p_3+p_4)p_5|1^+ \rangle -\br(1,2) \sq(2,5) t_{345} ]
\nonumber \\
&&+\frac{2\br(1,2) }{\br(1,3) \br(3,4) \br(4,5) \sq(1,2) \sq(2,5)
s_{15}s_{23}}[
\br(3,4) \sq(1,2) \sq(4,5) (2 \br(1,2) ^2 \sq(2,3) \sq(2,5) +\br(1,3)
\sq(3,5) \langle 1^-|p_4+p_5|3^- \rangle 
\nonumber \\
&&+\br(1,2) (\br(1,5) \sq(2,5) \sq(3,5) + \br(1,4) (\sq(2,5) \sq(3,4)
-\sq(2,3) \sq(4,5) )))-2\sq(2,5) s_{23}t_{345}\langle 1^-|p_2-p_3-p_4|5^-
\rangle ] \nonumber \\
&&-\frac{1}{4 \br(1,3) \br(1,4) \br(3,4) \sq(1,2) \sq(1,5) \sq(2,5) s_{23}s_{5{\sss H}}}
\left\{ 4 \br(1,3) ^2 \sq(3,5) ^3 s_{23} \langle 3^-|p_1+p_2|5^- \rangle \right. \nonumber \\
&&+\sq(3,5) \sq(4,5) \left[ \br(1,2) ^2 \br(3,4) \sq(2,5) ^2 (2 s_{23}-s_{12})-
2 \br(1,3) \sq(3,5) s_{23} (5 \br(1,4) \br(2,3) \sq(2,5) -\br(1,3)  \langle 4^-|6 p_1+p_2|5^- \rangle )
\right. \nonumber \\
&&\left. +s_{12} \br(1,3) ^2 \br(3,4) \sq(3,5) ^2 \right] +
\br(1,4) \sq(3,5) \sq(4,5) ^2 \left[ -\br(1,2) \br(3,4) \sq(2,5) s_{12}-
6 s_{23} (\br(1,4) \br(2,3) \sq(2,5) +\br(1,3) \langle 4^-|-2 p_1-p_2|5^- \rangle ) \right. \nonumber \\
&&\left. \left. +s_{12} \br(1,3) \br(3,4) \sq(3,5) \right] +
4 \br(1,4) ^2 \sq(4,5) ^3 s_{23} \langle 4^-| p_1+p_2|5^- \rangle \right\} \nonumber \\
&&-\frac{1}{4 \br(1,3) \br(1,4) \sq(1,2) \sq(2,5) s_{15} s_{23} s_{5{\sss H}} t_{123}} 
\left\{ \br(1,2) ^3 \sq(2,3) \sq(2,5) ^2 \sq(4,5)  s_{12} s_{45} \right. \nonumber \\
&&-4 \br(1,5) \br(2,3) \sq(2,5) \sq(3,4) \sq(3,5) (s_{13}+s_{23})  \langle 1^-| p_3+p_4|5^- \rangle ^2
+\br(1,2) ^3 \sq(2,5) \sq(4,5)  (2 s_{45} \sq(2,3)   \langle 2^+| p_3p_2+2p_1p_3|5^- \rangle
\nonumber \\
&& +\br(1,4) \sq(1,2) \sq(2,5) \langle 4^+| 4p_3p_5+p_5p_2|3^- \rangle +
\br(1,5) \sq(1,2) (-\br(2,4) \sq(2,5) ^2 \sq(3,4)
\nonumber \\
&& +\sq(3,5) (-s_{34} \sq(2,5) + \langle 5^+| -p_1p_4+p_4p_1|2^- \rangle ))) \nonumber  \\
&&+\br(1,2) ^2 \left[ 2 \br(1,3) \sq(1,2) \sq(3,5)  (2 \br(1,4) \sq(2,5) \sq(3,5) \sq(4,5) 
 \langle 5^-| p_2-p_3|4^- \rangle +\br(1,5) (2 \br(2,3) \sq(2,4) \sq(2,5) \sq(3,5) ^2
\right. \nonumber \\
&&+\sq(4,5) (-2 s_{34}\sq(2,5) \sq(3,5) +\br(1,4) \sq(1,5) \sq(2,3) \sq(4,5) +
\sq(2,5)  \langle 5^+|- p_2p_4+2p_4p_2|3^- \rangle ))) \nonumber \\
&&+\sq(4,5) (2 \br(1,4) \sq(2,5) ^2 s_{23} \langle 4^+|- p_5p_2-2p_3p_5|3^- \rangle +
\br(1,5) (-2\br(2,3) \sq(2,5)  (\br(2,4) \sq(2,3) \sq(2,5) ^2 \sq(3,4) \nonumber \\
&&+
\br(1,4) \sq(3,5) (\sq(1,4) \sq(2,3) \sq(2,5) -3 \sq(1,2) \sq(2,4) \sq(3,5) ))+
\br(1,4) \sq(1,2) \sq(4,5) (-3 s_{34} \sq(2,5) \sq(3,5) +\br(2,4) \sq(2,5) 
(-\sq(2,5) \sq(3,4) +3\sq(2,3) \sq(4,5) ) \nonumber \\
&&\left. +\br(1,4) (-\sq(1,4) \sq(2,5) \sq(3,5) +2 \sq(4,5) (\sq(1,5) \sq(2,3) +\sq(1,3) \sq(2,5) )
)))) \right] \nonumber \\
&&+\br(1,2) \left[ 2 \br(1,4) \br(1,5) \br(2,3) \sq(4,5) (-s_{23}\sq(2,5) \sq(3,5) 
(\sq(2,5) \sq(3,4) +2 \sq(2,4) \sq(3,5) ) +\sq(4,5) (-s_{34} \sq(2,3) \sq(2,5) \sq(3,5) 
\right. \nonumber \\
&&+\sq(2,3) \sq(2,5)  \langle 5^+|- p_2p_4+p_4p_2|3^- \rangle +\br(1,4) (-
\sq(1,4) \sq(2,3) \sq(2,5) \sq(3,5) +\sq(1,2) \sq(3,5) (\sq(2,5) \sq(3,4) +\sq(2,4) \sq(3,5) )
\nonumber \\
&&+2\sq(1,5) \sq(2,3) ^2 \sq(4,5) ))) + \br(1,3) ^2 \sq(3,5) ^2 (4 s_{25} 
\br(1,4) \sq(1,2) \sq(3,4) \sq(4,5) \nonumber \\
&&+\br(1,5) (4 \br(2,3) \sq(3,5)  (\sq(1,5) \sq(2,3) \sq(2,4) +\sq(1,2) (
\sq(2,5) \sq(3,4) +\sq(2,4) \sq(3,5) )) \nonumber \\
&&+\sq(1,2) \sq(4,5) (-s_{34} \sq(3,5) + \langle 5^+| p_1p_4-p_4p_1|3^- \rangle +
\langle 5^+| -3 p_2p_4+p_4p_2|3^- \rangle ))) \nonumber \\
&&+\br(1,3) \sq(3,5) (-4 \sq(1,4) \sq(2,5) \sq(3,4) \sq(4,5) s_{23} s_{25} +
\br(1,5) (-4 s_{23} \br(2,3) \sq(2,4) \sq(2,5) \sq(3,5) ^2 \nonumber \\
&&+\br(1,4) \sq(1,2) \sq(4,5) ^2 (-s_{34}  \sq(3,5) + \langle 5^+| p_1p_4-p_4p_1|3^- \rangle +
\langle 5^+| -3 p_2p_4+p_4p_2|3^- \rangle ) \nonumber \\
&&+2 \br(2,3) \sq(4,5) (\sq(2,3) \sq(2,5) \langle 3^+|-p_4p_3-p_2p_4|5^- \rangle
+\br(1,4) (\sq(3,5) (-\sq(1,4) \sq(2,3) \sq(2,5) \nonumber \\
&&\left. \left. +\sq(1,2) (5 \sq(2,5) \sq(3,4) +3 \sq(2,4) \sq(3,5) ))+
2 \sq(1,5) \sq(2,3) (\sq(2,4) \sq(3,5) +\sq(2,3) \sq(4,5) ))))) \right] \right\} \nonumber \\
&&-\frac{\langle 1^-|p_{\sss H}|5^- \rangle }{2 \br(1,3) \br(1,4) \br(3,4) \sq(1,5) \sq(2,5) s_{23}
s_{5{\sss H}} t_{234}} \left[
\br(1,2) \br(2,3) \br(3,4) \sq(3,5) ^2 (\br(1,3) \sq(3,4) \sq(2,5) +\br(1,4) \sq(2,4) \sq(4,5) )
\right. \nonumber \\
&&-2 \br(2,3) \sq(3,5) \langle 1^-|p_3+p_4|5^-\rangle ^2 (s_{23}+s_{34})+
\br(1,2) \br(1,3) \br(3,4) \sq(3,5) ^2 \langle 4^+|-p_2p_3-p_3p_4|5^- \rangle \nonumber \\
&&+\br(1,2) ^2 \br(3,4) \sq(2,5) \sq(3,5) \sq(4,5) (s_{23}+s_{24}) +
2 \br(1,3) \br(2,4) \sq(2,3) \sq(4,5) (\br(1,3) \br(2,3) \sq(3,5) ^2+\br(1,4) \br(2,4) \sq(4,5) ^2)
\nonumber \\
&&\left. +\br(1,2) \br(2,4) \br(3,4) \sq(3,5) \sq(4,5) ^2 \langle 1^-|p_4-p_3|2^-\rangle +
\br(1,4) \sq(3,5) \sq(4,5) ^2 (-4 s_{23} \br(1,3) \br(2,4) -s_{34} \br(1,2) \br(3,4) )\right]
\nonumber \\
&&+\frac{\br(1,2) \sq(3,5) \sq(4,5) }{4\br(1,3) \sq(1,2) \sq(2,5) s_{15}s_{23} s_{4{\sss H}}}
\langle 1^-|p_{\sss H}|4^- \rangle [(s_{12}-2 s_{23})\sq(2,5) +\br(1,3) \sq(1,2) \sq(3,5) ] \nonumber \\
&&-\frac{1}{4 \br(1,3) \br(1,4) \sq(1,2) \sq(2,5) s_{15} s_{23}s_{4{\sss H}} t_{123}}
\left\{ 4 \br(1,4) \br(2,3) \sq(2,5) \sq(3,5) ^2 (s_{13}+s_{23}) \langle 1^-|p_3+p_5|4^- \rangle ^2 
\right. \nonumber \\
&&-2 \br(1,2) ^3 \sq(2,5) \sq(4,5) s_{12} (-2 s_{45} \sq(2,3) \sq(2,5) +\br(1,4) \sq(1,2) 
(-\sq(2,5) \sq(3,4) +\sq(2,3) \sq(4,5) )) \nonumber \\
&&+\br(1,2) ^3 [ 4 \sq(2,5)  (\br(1,3) \sq(1,2) \sq(3,5) (2 s_{34} \sq(2,5) \sq(3,4) -
s_{45}\sq(2,3) \sq(4,5) )\nonumber \\
&&+ \sq(2,5) \sq(4,5) (s_{23}s_{45} \sq(2,3) -\br(1,5) \sq(1,2) 
\langle 3^+|p_4p_3+p_2p_4|5^- \rangle )) \nonumber  \\
&&-\br(1,4) \sq(1,2) (2 \br(2,3) \sq(2,5) (-2\sq(2,5) ^2 \sq(3,4) ^2-\sq(2,3) ^2\sq(4,5) ^2+
\sq(2,5) \sq(3,4) (2 \sq(2,4) \sq(3,5) +\sq(2,3) \sq(4,5) )) \nonumber \\
&&+\sq(4,5) (2\br(1,3) (2\sq(1,5) \sq(2,3) (\sq(2,5) \sq(3,4) +\sq(2,4) \sq(3,5) )+
\sq(3,5) (-2\sq(1,4) \sq(2,3) \sq(2,5) -\sq(1,3) \sq(2,4) \sq(2,5) \nonumber \\
&&+\sq(1,2) (\sq(2,5) \sq(3,4) +\sq(2,4) \sq(3,5) ))) +
\sq(2,5) (\sq(2,4) \sq(3,5) (3s_{35}-s_{25}) \nonumber \\
&&+\br(1,5) ( -2 \sq(1,2) \sq(3,5) \sq(4,5) +\sq(1,5) (\sq(2,5) \sq(3,4) -
3\sq(2,3) \sq(4,5) )))))] \nonumber \\
&&+\br(1,2) \sq(3,5) [2\br(1,3) ^3 \sq(1,2) \sq(3,4) \sq(3,5) \langle 5^+|2p_2p_4+p_4p_1|3^- \rangle 
\nonumber \\
&&+2 \br(1,4) \br(1,5) \br(2,3) \sq(2,5) \sq(4,5) (2\sq(2,3) \langle 5^+|-p_2p_3+p_3p_2|4^- \rangle
+(\sq(2,3) (s_{35}-s_{25})\nonumber \\
&&+\br(1,5) (\sq(1,5) \sq(2,3) -2\sq(1,2) \sq(3,5) ))\sq(4,5) ) 
+\br(1,3) ^2 \sq(3,5) (-4 \br(1,5) \br(2,4) \sq(1,2) \sq(2,5) \sq(3,4) \sq(4,5) \nonumber \\
&&+\br(1,4) (\sq(1,2) \sq(4,5) (\sq(3,4) (3s_{25}-s_{35})+
\br(1,5) (\sq(1,5) \sq(3,4) -2 \sq(1,3) \sq(4,5) )) \nonumber \\
&&-2 \br(2,3) \sq(3,4) (2\sq(1,5) \sq(2,3) \sq(2,4) +
\sq(1,2) (5\sq(2,5) \sq(3,4) +\sq(2,4) \sq(3,5) )))) \nonumber \\
&&-\br(1,3) (-4 \br(1,5)  \br(2,4) \sq(2,5) ^2 \sq(4,5) \sq(3,4) s_{23}+
\br(1,4) (-4 \br(2,3) \sq(2,4) \sq(2,5) \sq(3,4) \sq(3,5) s_{23} \nonumber \\
&&+\br(1,5) \sq(1,2) \sq(3,5) (3 s_{25}-s_{15}-s_{35}) \sq(4,5) ^2-
2\br(2,3) \sq(4,5)  ( \sq(2,5) \sq(2,3) \sq(3,4) (s_{25}-s_{35}) \nonumber \\
&&+\br(1,5) (\sq(3,5) (-2 \sq(1,4) \sq(2,3) \sq(2,5) +\sq(1,2) (5 \sq(2,5) \sq(3,4) +3 
\sq(2,4) \sq(3,5) ))+\sq(1,5) \sq(2,3) (3 \sq(2,4) \sq(3,5) +\sq(2,3) \sq(4,5) ))))) ] \nonumber \\
&&+\br(1,2) ^2 [4 \sq(2,5) (\br(1,3) ^2 \sq(1,2) \sq(3,4) ^2 \sq(3,5) 
 \langle 4^-|-2p_2+p_3|5^- \rangle -s_{23} \br(1,5) \sq(2,5) \sq(4,5) \langle 3^+|-p_4p_3-p_2p_4|5^- \rangle
\nonumber \\
&&+\br(1,3) \br(1,5) \sq(1,2) \sq(3,5) \sq(4,5) (-s_{34} \sq(3,5) +
\langle 5^+|-p_2p_4+p_4p_2|3^- \rangle )) +\br(1,4) (-4 s_{23}^2 \sq(2,5) ^2 \sq(3,4) \sq(4,5) \nonumber \\
&&+\sq(1,2) \sq(3,5) \sq(4,5) (2 \br(1,3) ^2 (-\sq(3,4) \sq(1,5) \sq(2,3) +\sq(1,3) \sq(2,4) \sq(3,5) )
+\br(1,5) \sq(2,5) (3s_{35} -s_{15} -s_{25}) \sq(4,5) \nonumber \\
&&+\br(1,3) (-s_{35} (3 \sq(2,5) \sq(3,4) +\sq(2,4) \sq(3,5) )+s_{25} (\sq(2,5) \sq(3,4) +3\sq(2,4) \sq(3,5) ) -3 s_{15} \sq(2,3) \sq(4,5) ))\nonumber \\
&&-2 \br(2,3) (\sq(2,5) \sq(4,5) (\sq(2,3) \sq(2,4) \sq(3,5) (s_{35}-s_{25})-
\br(1,5) (\sq(1,5) \sq(2,3) (-\sq(2,5) \sq(3,4) +\sq(2,3) \sq(4,5) )\nonumber \\
&&+\sq(1,2) \sq(3,5) (\sq(2,5) \sq(3,4) +3 \sq(2,4) \sq(3,5) )))+
\br(1,3) (2 \sq(1,5) \sq(2,3) ^2 \sq(2,5) \sq(3,4) \sq(4,5) \nonumber \\
&&\left. +\sq(1,2) \sq(3,5) (2 \sq(2,5) ^2 \sq(3,4) ^2 -\sq(2,3) \sq(2,4) \sq(3,5) \sq(4,5) +
\sq(2,5) \sq(3,4) (4\sq(2,4) \sq(3,5) +\sq(2,3) \sq(4,5) )))))] \right\} \nonumber \\
&&+\frac{\br(1,2) }{\br(1,3) \sq(1,2) s_{15} s_{4{\sss H}}t_{125}}[-\langle 4^+|p_3(p_1+p_2)|5^- \rangle -
\sq(4,5) t_{125}][\br(1,3) \sq(3,4) \sq(3,5) +\langle 1^-|p_2+p_5|3^- \rangle \sq(4,5) ] \nonumber \\
&&-\frac{1}{\br(1,4) \br(4,5) \sq(1,2) \sq(2,5) s_{15}s_{3{\sss H}}}[
\br(1,2) (\br(1,4) \sq(1,5) \sq(2,4) -\sq(2,5) (s_{12}+s_{15}+s_{24}+s_{25}))\sq(3,5)  
\langle 1^-|-p_4-p_5|3^- \rangle \nonumber \\
&&-\sq(4,5) (\br(1,4) ^2 \sq(3,4) ^2 \langle 4^-|-p_1-p_2|5^- \rangle -
\sq(3,5) (\br(1,2) ^2 \br(4,5) \sq(2,3) \sq(2,5) -\br(1,4) \br(1,5) \br(2,4) \sq(2,5) \sq(3,4) 
\nonumber \\
&&-\sq(3,4) \br(1,4) ^2 (-2s_{15}-s_{25})))+\br(1,5) \sq(3,5) ^2 (s_{15}+s_{25})\br(1,4) \sq(4,5) ] 
\nonumber  \\
&&-\frac{\br(1,2) }{\br(1,4) \sq(1,2) s_{15} s_{3{\sss H}}t_{125}}[-\langle 3^+|p_4(p_1+p_2)|5^- \rangle -
\sq(3,5) t_{125}][\br(1,2) \sq(2,4) \sq(3,5) +\langle 1^-|p_4+p_5|3^- \rangle \sq(4,5) ] \nonumber \\
&&+\frac{\langle 1^-|p_4+p_5|3^- \rangle \sq(4,5) }{\br(1,3) \br(4,5) \sq(2,5) s_{15} s_{3{\sss H}} t_{145}}
[ \br(1,2) (-s_{25} \br(1,3) \sq(3,5) +\br(1,3) (-\langle 3^+|p_4(p_1+p_2)|5^- \rangle
-\sq(3,5) (s_{12}+s_{15}) ))\nonumber \\
&&-\br(1,3) \br(2,4) \sq(4,5) \langle 1^-|p_{\sss H}|3^- \rangle ] \nonumber \\
&&+\frac{\br(1,2) ^2}{\br(1,3) \br(3,4) \br(4,5) s_{15}} \langle 1^-|p_3+p_4|5^- \rangle \nonumber \\
&& -\frac{\br(1,2) ^2\sq(4,5) }{\br(1,3)  \br(4,5) s_{15}t_{145}} 
\langle 1^-|p_4+p_5|3^- \rangle \nonumber \\
&&+\frac{\br(1,2) ^3  \langle 1^-|p_{\sss H}|2^- \rangle }{\br(1,3) \br(1,5) \br(3,4) \br(4,5) s_{1{\sss H}}} \nonumber \\
&&+\frac{\br(1,2) ^3  }{\br(1,3) \br(1,4) \br(1,5) \br(2,3) \br(4,5) s_{1{\sss H}}} 
\langle 1^-|(p_4+p_5)(p_2+p_3)|1^+ \rangle \nonumber \\
&&\left.-\frac{\br(1,2) ^3  \langle 1^-|p_{\sss H}|5^- \rangle }{\br(1,4) \br(1,5) \br(2,3) \br(3,4) s_{1{\sss H}}}
\right\}  \, . 
\label{sec:Ammppp}
\end{eqnarray} 
All the other sub-amplitudes can be obtained by relabelling and
by use of reflection symmetry, and parity inversion.

\subsection{Sub-amplitudes for Higgs $+$ three gluons and a $q\qb$ pair}
\label{sec:appbb}

The colour decomposition of the amplitudes for Higgs, three gluons
and a $q\bar q$ pair is given in \eqn{TwoQuarkGluonDecomp}, with $n$ = 5.
There are four independent sub-amplitudes, which we may take to be the
one which has no gluons of negative helicity, and the three with one gluon
of negative helicity in different positions. They are
\begin{eqnarray}
\lefteqn{m_5(1_q^+,2_{\bar{q}}^-;3^+,4^+,5^+)=i A  
\left\{ -\frac{(s_{12}+s_{13})\langle 2^-|(p_1+p_3)p_{\sss H}|2^+\rangle ^2}
{s_{12} \br(1,3) \br(2,3) \br(2,4) \br(2,5) \br(4,5) t_{123}}
\right. } \nonumber \\
&& -\frac{\sq(1,5) \langle 2^-|(p_1+p_5)p_{\sss H}|2^+\rangle ^2}
{s_{12}  \br(2,3) \br(2,4) \br(2,5) \br(3,4) t_{125}}-
\frac{ \langle 2^-|p_{\sss H}|1^- \rangle^2}
{ \br(2,3) \br(2,5) \br(3,4) \br(4,5) \sq(1,2) } \nonumber \\
&&+\frac{ \langle 2^-|p_{\sss H}|5^- \rangle^2}
{s_{12} \br(1,3) \br(2,3) \br(2,4) \br(3,4)   s_{5{\sss H}}}
\left[- \frac{(s_{12}+s_{13}) \langle
2^-|(p_1+p_3)p_4|3^+\rangle}{t_{123}}
\right. \nonumber \\
&& \left. -\langle 2^-|(p_3+p_4)p_1|3^+\rangle 
  -s_{12} \br(2,3) \right] \nonumber
\\
&& +\frac{ \langle 2^-|p_{\sss H}|4^- \rangle^2}
{s_{12} \br(1,3) \br(2,3) \br(2,5)  s_{4{\sss H}}}\left[ \frac{\sq(1,5) 
 \langle 1^-|p_3 (p_1+p_5) |2^+ \rangle }{t_{125}}-
\frac{(s_{12}+s_{13}) \langle 2^-|p_1+p_3|5^- \rangle}{t_{123}} \right]
\nonumber \\
&& \left. +\frac{ \langle 2^-|p_{\sss H}|3^- \rangle^2}
{s_{12} \br(2,4) \br(2,5) \br(4,5)   s_{3{\sss H}}}\left[ 
\langle 2^-|p_4+p_5|1^- \rangle -\frac{\br(4,5) \sq(1,5) \langle
2^-|p_1+p_5|4^- \rangle}
{t_{125}} \right] \right\} \, ,
\label{sec:Aqqppp}
\end{eqnarray}

\begin{eqnarray}
\lefteqn{m_5(1_q^+,2_{\bar{q}}^-;3^-,4^+,5^+)=i A 
\left\{
\frac{\br(2,3) \sq(4,5) ^2}{s_{34}s_{12}s_{45}} \langle 3^-|p_4+p_5-p_2 
|1^- \rangle \right. } 
\nonumber \\
&&+\frac{\br(2,3) \sq(4,5) }{2 \br(3,5) s_{34}s_{12}t_{123}} \left[
\sq(1,4) \langle 3^-|p_1 (p_4+p_5)+p_2(p_4-p_5)|3^+ \rangle
\right. \nonumber\\
&&\left. +2 \br(2,3) ( \br(3,5) \sq(1,5) \sq(2,4) -\sq(1,2) \langle
3^-|p_1+p_2|4^- \rangle)
\right] \nonumber \\
&&+\frac{\br(2,3) \sq(1,4) \sq(4,5) }{\br(3,5) \sq(1,3) s_{34}
t_{123}} \langle 3^-|p_4+p_5-p_2 |1^- \rangle \nonumber \\
&&-\frac{\sq(4,5) }{\br(3,5) s_{34} s_{12}s_{45}t_{123}}
\left[ \langle 3^-|p_1+p_2 |4^- \rangle (\br(1,2) \langle 3^-|p_4+p_5 |1^-
\rangle ^2- \br(2,3) ^2 \sq(1,2) s_{45})\right. \nonumber \\
&& \left. +\langle 3^-|p_4+p_5 |1^-
\rangle \br(2,3) \br(3,5) \sq(4,5) (s_{13}+s_{23}) \right] \nonumber
\\
&&+\frac{\br(1,3) \sq(1,4) \sq(4,5) }{\br(3,5)
s_{34}s_{13}s_{45}t_{123}}
\left[\br(1,2) \langle 3^-|p_4+p_5 |1^- \rangle ^2-\br(2,3)
(s_{34}+s_{35})
\langle 3^-|p_4+p_5 |1^- \rangle -s_{45}\br(2,3) ^2 \sq(1,2) \right]
\nonumber \\
&&-\frac{\sq(4,5) ^2}{s_{34}s_{12}s_{45}t_{345}} \left[
\br(1,2) \langle 3^-|p_4+p_5 |1^- \rangle ^2-\br(2,3) ^2 \sq(1,2)
t_{345} \right] \nonumber \\
&&+\frac{\br(2,3) \sq(1,4) \sq(4,5) }{2 s_{34}s_{12}s_{5{\sss H}}} \langle
3^-|p_{\sss H}|5^- \rangle \nonumber \\
&&-\frac{1}{2 \br(3,5) s_{34}s_{12}t_{123}s_{5{\sss H}}} \left[
2 \br(1,2) \br(3,5) \sq(1,5) ^2 \langle 3^-|p_1+p_2|4^- \rangle^2
-2 \br(2,3) \sq(1,4) \sq(4,5) \langle 3^-|(p_1p_3-p_4p_1)p_5p_1|3^+ \rangle 
 \right.
\nonumber \\
&&+\br(1,3) \br(2,3) \br(3,5) \sq(1,4) \sq(1,5) \sq(4,5)
(s_{24}-s_{12}-s_{14})-
2 \br(1,3) \br(2,5) \sq(1,4) \sq(1,5) \sq(4,5) \langle
3^-|(p_1+p_2)p_4|3^+ \rangle \nonumber \\
&& +\br(2,3)  \br(3,5) \sq(4,5) ^2 \langle
3^-|p_4 p_2(p_3+p_4)|1^- \rangle +\br(2,3) \br(3,5) \sq(4,5) ^2 
\langle 1^+|(p_3p_2+p_4p_3)p_4|3^+ \rangle \nonumber \\
&&+2 \br(2,3) \sq(1,4) \sq(4,5) \langle 3^-|p_2p_5(p_3+p_4)p_1|3^+\rangle \nonumber
+2 \br(2,3) ^2 \br(3,5) \sq(1,5) \sq(2,4) \sq(4,5) (s_{13}+s_{23})
\nonumber \\
&&+\br(2,3) ^2 \br(3,5) \sq(1,4) \sq(2,5) \sq(4,5)
(s_{24}+s_{34}-s_{12}-s_{14}) -2\br(1,3) \br(2,3) ^2 \sq(1,2) \sq(1,4)
\sq(4,5) (s_{35}+s_{45})\nonumber \\
&&-\br(3,4) \br(3,5) \sq(1,4) \sq(4,5) ^2 \langle 3^-|(p_2+p_4)p_1|2^+
\rangle \nonumber \\
&&+\br(1,3) \br(3,4) \br(3,5) \sq(1,4) \sq(4,5) (\br(2,4) \sq(1,4)
\sq(4,5) -\br(2,3) \sq(1,5) \sq(3,4) ) \nonumber \\
&&+2 \br(1,3) \br(3,5) \sq(1,4) \sq(4,5) (-\sq(1,5) \langle
3^-|p_1p_4|2^+ \rangle + \br(2,3) \langle 1^+|p_3p_4|5^- \rangle
)\nonumber \\
&&\left. +2 \br(2,3) \br(3,5) \sq(2,4) \sq(4,5) (-\sq(1,5) \langle
2^-|p_1p_4|3^+ \rangle + \br(2,3) \langle 1^+|p_2p_4|5^- \rangle 
) \right] \nonumber \\
&&+\frac{\br(1,3) \sq(1,4) }{2 s_{34}s_{13}t_{123}s_{5{\sss H}}} \left[ 
2 \langle 2^-|p_{\sss H}|5^- \rangle (\br(3,4) \sq(1,4) \sq(4,5) +\sq(1,5)
\langle 3^-|p_1+p_2|4^- \rangle ) +\langle 3^-|p_{\sss H}|5^- \rangle \br(2,3)
\sq(1,3) \sq(4,5) \right] \nonumber \\
&&+\frac{\br(1,3) \sq(1,4) ^2}{s_{34}s_{13}t_{134}s_{5{\sss H}}} \langle
2^-|p_{\sss H}|5^- \rangle  \langle 3^-|p_1+p_4|5^- \rangle  -
\frac{\br(2,3) \sq(1,4) \sq(4,5) }{2 \br(3,5) \sq(3,4) s_{12} s_{4{\sss H}}}
\langle 3^-|p_{\sss H}|4^- \rangle \nonumber \\
&&-\frac{\br(1,3) \br(2,3) \sq(1,4) ^2 \sq(2,5) }{\br(3,5) \sq(3,4)
s_{13}s_{25}s_{4{\sss H}}} \langle 2^-|p_{\sss H}|4^- \rangle
\nonumber \\
&&-\frac{\sq(3,5) }{2 s_{34}s_{12}s_{35}t_{123}s_{4{\sss H}}} \left[
\langle 3^-|-p_1-p_2|4^- \rangle \br(3,4) \sq(1,4) (2 \br(1,2) \br(1,3)
\sq(1,4) \sq(1,5) +\br(1,2) \br(2,3) \sq(1,5) \sq(2,4)
\right. \nonumber \\
&&+\br(1,2) \br(2,3) \sq(1,4) \sq(2,5) -\br(1,5) \br(2,3) \sq(1,5)
\sq(4,5) +
2 \br(1,3) \br(2,5) \sq(1,5) \sq(4,5) -s_{35} \br(2,3)  \sq(4,5) )
\nonumber \\
&&-2 \br(2,3) \sq(1,4) \sq(4,5) \langle 3^-|(p_1p_5p_2-p_2p_3p_5)p_4|3^+ 
\rangle + \br(2,3) \sq(1,4) \sq(4,5) \langle
3^-|(p_1p_4+p_4p_2)p_5p_2|3^+ \rangle \nonumber \\
&&-\br(3,4) \br(3,5) \sq(1,4) \sq(4,5) ^2  \langle 3^-|(p_2+p_5)p_1|2^+
\rangle +
\br(2,3) \br(3,5) \sq(4,5) ^2 \langle 1^+|p_5(p_2+p_3)p_4|3^+\rangle
\nonumber \\
&&-\br(2,3) \br(3,5) \sq(4,5) ^2 \langle 3^-|p_4p_5(p_2+p_3)|1^- \rangle 
+2 \br(2,3) ^2 \sq(1,4) \sq(4,5) (s_{45}\br(1,3) \sq(1,2) -\langle
3^-|p_5p_1p_4|2^-\rangle )
\nonumber \\
&& +\br(1,3) \br(3,5) \sq(1,4) \sq(4,5) (\br(2,3) \langle
1^+|-p_4p_2|5^-\rangle +
\sq(1,5) \langle 2^-|p_5p_4|3^-\rangle ) \nonumber \\
&&\left. +2 \br(1,3) \br(3,4) \br(3,5) \sq(1,4) \sq(4,5) (\br(1,2) \sq(1,4)
\sq(1,5) -\br(2,3) \sq(1,3) \sq(4,5) ) \right] \nonumber \\
&&-\frac{\br(1,3) \sq(1,4) }{2\br(3,5) \sq(3,4) s_{13}t_{123}s_{4{\sss H}}}
\left[ 2\langle 3^-|-p_1-p_2|5^- \rangle \sq(1,4) (\langle
2^-|-p_1-p_3|4^- \rangle +
\br(2,5) \sq(4,5) ) \right. \nonumber \\
&&\left. +\br(2,3) \sq(1,3) \sq(4,5) \langle 3^-|p_{\sss H}|4^- \rangle +
2 \br(3,5) \sq(1,5) \sq(4,5) \langle 2^-|p_{\sss H}|4^- \rangle \right]
\nonumber \\
&&+\frac{1}{\br(3,5) \sq(3,4) s_{12} t_{125} s_{4{\sss H}}} \langle
2^-|p_1+p_5|4^- \rangle \langle
3^-|p_{\sss H}|4^- \rangle \left[ 
\br(2,3) \sq(1,4) \sq(2,5) -\sq(1,5)   \langle
3^-|p_1+p_5|4^- \rangle \right] \nonumber \\
&&+\frac{\br(2,3) \sq(1,4) \sq(2,5) }{\br(3,5) \sq(3,4) s_{25} t_{125} s_{4{\sss H}}} \langle
2^-|p_1+p_5|4^- \rangle \langle
3^-|p_{\sss H}|4^- \rangle \nonumber \\
&&+\frac{\br(2,3) ^2 \sq(1,2) \sq(4,5) }{\br(3,4) \br(3,5) s_{12}
s_{45}s_{3{\sss H}}} \langle 3^-|(p_1+p_2)(p_4+p_5)|3^+\rangle \nonumber \\
&&-\frac{\br(2,3) ^2 \sq(1,2) }{\br(3,4) \br(3,5) s_{12}t_{125}s_{3{\sss H}}}\langle
3^-|p_1+p_2|5^- \rangle \langle
3^-|p_{\sss H}|4^- \rangle \nonumber \\
&&+\frac{\br(2,3) ^2 \sq(2,5) }{\br(3,4) \br(3,5) s_{25}t_{125}s_{3{\sss H}}}\langle
3^-|p_2+p_5|1^- \rangle \langle
3^-|p_{\sss H}|4^- \rangle \nonumber \\
&&+\frac{\br(2,3) ^2 \sq(2,5) }{\br(3,4) \br(3,5) s_{25}t_{245}s_{3{\sss H}}}\langle
3^-|p_2+p_5|4^- \rangle \langle
3^-|p_{\sss H}|1^- \rangle \nonumber \\
&&+\frac{\br(2,3) ^2 \sq(4,5) }{\br(3,4) \br(3,5) s_{45}t_{245}s_{3{\sss H}}}\langle
3^-|p_4+p_5|2^- \rangle \langle
3^-|p_{\sss H}|1^- \rangle \, ,
\label{sec:Aqqmpp}
\end{eqnarray}

\begin{eqnarray}
\lefteqn{m_5(1_q^+,2_{\bar{q}}^-;3^+,4^-,5^+)=i A  \left\{
\frac{\br(2,4) \sq(3,5) ^2}{s_{45}s_{12}s_{34}} \langle
4^-|p_3+p_5-p_2|1^-\rangle \right. }\nonumber \\
&&+\frac{\br(2,4) \sq(3,5) }{2\br(3,4) s_{45}s_{12}t_{125}} \left[
\langle 4^-|p_1+p_2|5^- \rangle \langle 4^-|p_3-2 p_2|1^- \rangle -
\langle 4^-|p_2-p_5|1^- \rangle \br(3,4) \sq(3,5) \right] \nonumber  \\
&&+\frac{\br(2,4) ^2 \sq(2,5) \sq(3,5) }{\br(3,4) s_{45}s_{25}t_{125}}
\langle 4^-|p_2-p_3+p_5|1^- \rangle \nonumber \\
&&+\frac{\sq(3,5) }{s_{45}s_{12}s_{34}t_{125}} \left[
-\langle 4^-|p_1+p_2|5^- \rangle (\br(1,2) \br(3,4) \sq(1,3) ^2+\br(2,4) 
^2 \sq(1,2) \sq(3,4) )\right. \nonumber \\
&&\left. +\br(3,4) \sq(1,3) \sq(1,5) \sq(3,5) (\br(1,4) \br(2,5)
+\br(1,2) \br(4,5) ) +\br(2,5) \br(3,4) \sq(3,5) (\br(2,4) \sq(1,3)
\sq(2,5) + \br(4,5) \sq(1,5) \sq(3,5) )\right] \nonumber  \\
&&+\frac{\br(2,4) \sq(2,5) \sq(3,5) }{s_{45} s_{25}s_{34}t_{125}}
\left[ \sq(1,3) \langle 2^-|(-p_1-p_5)p_3|4^+ \rangle +\br(2,4) \langle
3^+|p_4(p_2+p_5)|1^- \rangle \right] \nonumber \\
&&-\frac{\sq(3,5) ^2}{s_{45}s_{12}s_{34}t_{345}} \left[
\br(1,2) \langle 4^-|p_3+p_5|1^- \rangle ^2-\br(2,4) ^2 \sq(1,2) t_{345} 
\right] \nonumber \\
&&+\frac{\br(2,4) \sq(1,5) \sq(3,5) }{\br(3,4) \sq(4,5) s_{12}s_{5{\sss H}}}
\langle 4^-|p_{\sss H}|5^- \rangle \nonumber \\
&&+\frac{\sq(3,5) }{\br(4,5) s_{12}s_{34}s_{5{\sss H}}} \left[
\br(1,2) \sq(1,5) ^2 \langle 4^-|p_1+p_2|3^- \rangle +
\sq(1,3) \sq(1,5) \sq(3,5) (\br(1,4) \br(2,3) -\br(1,2) \br(3,4)
)\right. \nonumber \\
&&\left. -\br(2,4) \sq(1,5) \sq(3,5) (s_{14}+s_{23}+s_{24})-
\br(3,4) \sq(3,5) ^2 \langle 2^-|p_3+p_4|1^- \rangle \right] \nonumber
\\
&&+\frac{1}{\br(3,4) \sq(4,5) s_{12}t_{123} s_{5{\sss H}}}
\langle 2^-|p_1+p_3|5^- \rangle \langle 4^-|p_{\sss H}|5^- \rangle \left[
\sq(1,3) \langle 4^-|-p_1-p_3|5^- \rangle +\br(2,4) \sq(1,5) \sq(2,3)
\right] \nonumber \\
&&-\frac{\sq(1,3) }{\br(3,4) \sq(4,5) s_{13}t_{123}s_{5{\sss H}}}
\langle 2^-|p_1+p_3|5^- \rangle \langle 4^-|p_1+p_3|5^- \rangle
\langle 4^-|p_{\sss H}|5^- \rangle \nonumber \\
&&-\frac{\sq(1,3) }{\br(3,4) \sq(4,5) s_{13}t_{134}s_{5{\sss H}}}
\langle 4^-|p_1+p_3|5^- \rangle ^2
\langle 2^-|p_{\sss H}|5^- \rangle \nonumber \\
&&+\frac{\sq(1,3) \sq(3,5) }{ \sq(4,5) s_{34}t_{134}s_{5{\sss H}}}
\langle 4^-|p_1+p_3|5^- \rangle 
\langle 2^-|p_{\sss H}|5^- \rangle \nonumber \\
&&+\frac{\br(2,4) ^2 \sq(1,3) \sq(2,5) }{\br(3,4) \br(4,5) s_{13}s_{25}s_{4{\sss H}}}
\langle 4^-|(p_2+p_5)(p_1+p_3)|4^+ \rangle \nonumber \\
&&-\frac{\br(2,4) ^2 \sq(1,2) }{\br(3,4) \br(4,5) s_{12}t_{123}s_{4{\sss H}}}
\langle 4^-|p_1+p_2|3^- \rangle \langle 4^-|p_{\sss H}|5^- \rangle \nonumber \\
&&-\frac{\br(2,4) ^2 \sq(1,3) }{\br(3,4) \br(4,5) s_{13}t_{123}s_{4{\sss H}}}
\langle 4^-|p_1+p_3|2^- \rangle \langle 4^-|p_{\sss H}|5^- \rangle \nonumber \\
&&-\frac{\br(2,4) ^2 \sq(1,2) }{\br(3,4) \br(4,5) s_{12}t_{125}s_{4{\sss H}}}
\langle 4^-|p_1+p_2|5^- \rangle \langle 4^-|p_{\sss H}|3^- \rangle \nonumber \\
&&+\frac{\br(2,4) ^2 \sq(2,5) }{\br(3,4) \br(4,5) s_{25}t_{125}s_{4{\sss H}}}
\langle 4^-|p_2+p_5|1^- \rangle \langle 4^-|p_{\sss H}|3^- \rangle \nonumber \\
&&-\frac{\br(2,4) \sq(1,5) \sq(3,5) }{2 s_{45}s_{12}s_{3{\sss H}}}
\langle 4^-|p_{\sss H}|3^- \rangle \nonumber \\
&&-\frac{1}{2\br(3,4) s_{45}s_{12}t_{125}s_{3{\sss H}}} \left[
2 \br(1,2) \br(3,4) \sq(1,3) ^2  \langle 4^-|p_1+p_2|5^- \rangle ^2 \right. \nonumber \\
&&- \br(2,4) \br(3,4)  \sq(1,5) \sq(3,5) 
\langle 4^-|p_1+p_5|3^- \rangle (s_{12}-s_{14}) 
 \nonumber \\
&&-2 (\br(2,5) \br(3,4) -\br(2,3) \br(4,5) )\br(1,4) \sq(1,3) \sq(1,5) \sq(3,5) 
\langle 4^-|p_1+p_2|5^- \rangle \nonumber \\
&&-2\langle 4^-|p_1+p_2|5^- \rangle (-s_{13} \br(2,4)  +
\br(2,5) \br(3,4) \sq(3,5) ) \br(4,5) \sq(1,5) \sq(3,5) \nonumber \\
&&-\br(2,4) \sq(1,5) \sq(3,5) \langle 4^-|p_2p_1(p_2+p_4)p_3|4^+ \rangle +
2 \br(2,4) \br(3,4) \sq(1,3) \sq(3,5) 
\langle 5^+|p_2(p_4+p_5)p_1|4^+ \rangle \nonumber \\
&&-\br(2,4) ^2 \br(3,4) \sq(1,3) \sq(2,5) \sq(3,5)  (s_{24}+2 s_{25}) \nonumber \\
&&-\br(2,4) ^2 \br(3,4) \sq(3,5) ^2 \langle 1^+|p_4p_5+p_2p_4|2^- \rangle \nonumber \\
&&\left. +\br(2,4) \br(3,4) \br(4,5) \sq(3,5) \sq(4,5) (\br(1,4) \sq(1,3) \sq(1,5) +
\br(2,4) \sq(1,3) \sq(2,5) +\br(4,5) \sq(1,5) \sq(3,5) )  \right] \nonumber \\
&&-\frac{\br(2,4) \sq(2,5) }{2 s_{45}s_{25}t_{125}s_{3{\sss H}}} \left[
-2 \br(1,2) \sq(1,3) ^2 \langle 4^-|p_1+p_2|5^- \rangle +\br(2,5) \sq(1,5) \sq(3,5) 
\langle 4^-|p_{\sss H}|3^- \rangle \right. \nonumber \\
&&\left. -2 \sq(1,3) \sq(3,5) (\langle 2^-|(p_4+p_5)p_1|4^+ \rangle +\br(2,4) t_{245}) \right]
\nonumber \\
&&\left. -\frac{\br(2,4) ^2 \sq(1,3) \sq(2,5) }{s_{45} s_{25}t_{245}s_{3{\sss H}}}
[-\langle 3^+|p_1(p_2+p_4)|5^- \rangle -\sq(3,5) t_{245}] \right\} \, ,
\label{sec:Aqqpmp}
\end{eqnarray}

\begin{eqnarray}
\lefteqn{m_5(1_q^+,2_{\bar{q}}^-;3^+,4^+,5^-)= i A  \left\{
\frac{\br(2,5) \sq(3,4) ^2}{s_{12}s_{34}s_{45} }
\langle 5^-|p_3+p_4-p_2 |1^-\rangle  \right. }\nonumber \\
&&+\frac{\br(2,5) \sq(3,4) }{2 \br(3,5) s_{12} s_{45}t_{125}}
\left[ \langle 5^-|2 p_2-p_3-p_4 |1^-\rangle  \langle 5^-|-p_1-p_2 |4^-\rangle -
\br(2,5) \br(3,5) \sq(1,2) \sq(3,4)
\right] \nonumber \\
&&+\frac{\br(2,5) ^2 \sq(2,4) \sq(3,4) }{\br(3,5) s_{25} s_{45}
t_{125}}\langle 5^-|p_2-p_3-p_4 |1^-\rangle
\nonumber \\
&&+\frac{\sq(3,4) }{\br(3,5) s_{12}s_{34}s_{45}t_{125}}\left[
\br(1,2) \br(1,5) \sq(1,4) \langle 5^-|p_3+p_4|1^-\rangle ^2
\right. \nonumber \\
&& +\br(2,5) \br(3,5) \sq(1,3) \sq(2,4) ( \langle 2^-|-p_1p_3|5^+\rangle
-\br(2,5) s_{14} )  +\br(2,5) ^2 \sq(1,4) 
(\langle 5^-|p_1p_4p_3|2^-\rangle   -s_{13}\langle 5^-|p_4|2^-\rangle
)\nonumber \\
&&-\br(2,5) \br(3,5) \sq(3,4) (s_{15}+s_{25})\langle
5^-|p_3+p_4|1^-\rangle
+\br(2,5) \br(4,5) \sq(1,4) \sq(2,4) \langle
5^-|-2p_1p_3+p_4p_1|2^+\rangle \nonumber \\
&&\left. -\br(2,5) ^3 \sq(1,2) \sq(2,4) s_{34} \right] \nonumber \\
&&+\frac{\br(2,5) \sq(3,4) }{\br(3,5) s_{25}s_{34}s_{45}t_{125}}
\left[  \langle 5^-|p_3+p_4|1^-\rangle ^2 \langle 1^-|p_2+p_5|4^-\rangle
-\br(2,5) ^2 \sq(1,2) \sq(2,4) s_{34}\right. \nonumber \\
&&\left. -s_{25} \br(3,5) \sq(3,4)
\langle 5^-|p_3+p_4|1^-\rangle \right] \nonumber \\
&&-\frac{\sq(3,4) ^2}{s_{12}s_{34}s_{45} t_{345}}
\left[ \br(1,2)  \langle 5^-|p_3+p_4|1^-\rangle ^2 -t_{345}\br(2,5) ^2
\sq(1,2) \right] \nonumber \\
&&-\frac{\br(2,5) ^2 \sq(1,2) \sq(3,4) }{\br(3,5) \br(4,5)
s_{12}s_{34}s_{5{\sss H}}}
 \langle 5^-|(p_1+p_2)(p_3+p_4)|5^+\rangle \nonumber \\
&&-\frac{\br(2,5) ^2 \sq(1,2)  }{\br(3,5) \br(4,5)
s_{12}s_{5{\sss H}}t_{123}}
 \langle 5^-|p_1+p_2|3^-\rangle \langle 5^-|p_{\sss H}|4^-\rangle \nonumber \\
&&-\frac{\br(2,5) ^2 \sq(1,3)  }{\br(3,5) \br(4,5)
s_{13}s_{5{\sss H}}t_{123}}
 \langle 5^-|p_1+p_3|2^-\rangle \langle 5^-|p_{\sss H}|4^-\rangle \nonumber \\
&&-\frac{\br(2,5) ^2 \sq(1,3)  }{\br(3,5) \br(4,5)
s_{13}s_{5{\sss H}}t_{134}}
 \langle 5^-|p_1+p_3|4^-\rangle \langle 5^-|p_{\sss H}|2^-\rangle \nonumber \\
&&+\frac{\br(2,5) ^2 \sq(3,4)  }{\br(3,5) \br(4,5)
s_{34}s_{5{\sss H}}t_{134}}
 \langle 5^-|p_3+p_4|1^-\rangle \langle 5^-|p_{\sss H}|2^-\rangle \nonumber \\
&&+\frac{\br(2,5) \sq(1,4)  \sq(3,4)  }{2\br(3,5) \sq(4,5)
s_{12}s_{4{\sss H}}}
 \langle 5^-|p_{\sss H}|4^-\rangle \nonumber \\
&&-\frac{\br(2,5)  \sq(1,3)  }{\br(3,5) \sq(4,5)
s_{13}s_{25}s_{4{\sss H}}}
 \langle 5^-|p_1+p_3|4^-\rangle \langle 4^+|(p_1+p_3)(p_2+p_5)|4^-\rangle
\nonumber \\
&&+\frac{1  }{\br(3,5) \sq(4,5)
s_{12}s_{4{\sss H}}t_{123}}
 \langle 2^-|p_1+p_3|4^-\rangle \langle 5^-|p_{\sss H}|4^-\rangle \left[ -\sq(1,4)
 \langle 5^-|p_1+p_2|3^-\rangle +\br(3,5) \sq(1,3) \sq(3,4) \right]
\nonumber \\
&&-\frac{  \sq(1,3)  }{\br(3,5) \sq(4,5)
s_{13}t_{123}s_{4{\sss H}}}
 \langle 2^-|p_1+p_3|4^-\rangle \langle 5^-|p_1+p_3|4^-\rangle \langle 5^-|p_{\sss H}|4^-\rangle
\nonumber \\
&&-\frac{ 1  }{2\br(3,5) s_{45}
s_{12}t_{125}s_{4{\sss H}}}\left[
\br(4,5) \sq(1,4) \langle 5^-|p_1+p_2|4^-\rangle (2 \br(1,2) \br(1,5)
\sq(1,3) \sq(1,4) \right. \nonumber \\
&&+\br(1,2) \br(2,5) \sq(1,4) \sq(2,3) +\br(1,2) \br(2,5) \sq(1,3)
\sq(2,4) -2\br(1,5) \br(2,3) \sq(1,3) \sq(3,4) -s_{13}\br(2,5)
\sq(3,4) )\nonumber \\
&&+\br(2,5) \sq(1,4) \sq(3,4)  \langle
5^-|p_1(p_4p_2p_3+p_3p_4p_2)|5^+\rangle -
2 \br(1,5) \br(2,3) \sq(1,3) \sq(1,4)  \sq(3,4) \langle 5^-|(p_2+p_3)p_4|5^+\rangle \nonumber \\
&&+\br(2,5) ^2 \sq(1,4) \sq(3,4) (-s_{23}\br(4,5) \sq(2,4) -s_{34}\br(1,5) \sq(1,2) )+
\br(2,5) \br(3,5) \br(4,5) \sq(1,4) \sq(3,4) ^2 (s_{12}+s_{13}+s_{15}) \nonumber \\
&&+\br(2,5) \br(3,5) \br(4,5) \sq(3,4) ^2 (s_{25}\sq(1,4) -\langle 1^+|p_3p_2|4^- \rangle )+
\br(2,5) \br(3,5) \br(4,5) \sq(3,4) ^3 \langle 3^-|p_2+p_5|1^- \rangle \nonumber \\
&&+s_{45} \br(2,5) \sq(3,4) \sq(1,4) \langle 5^-|(p_1+p_2)p_3|5^+\rangle \nonumber \\
&&\left. -2 \br(3,5) \sq(1,3) \sq(1,4) \sq(3,4) (\br(1,2) \br(1,5) \br(4,5)
\sq(1,4) +\br(1,4) \br(2,5) ^2 \sq(2,4) )
-s_{45}\br(2,5) \br(3,5) ^2 \sq(1,3) \sq(3,4) ^2 \right] \nonumber \\
&&+\frac{\br(2,5) }{2 \br(3,5) \sq(4,5) s_{25} s_{4{\sss H}}t_{125}} \left[
2 \br(1,2) \sq(1,4) ^2 \sq(2,4) \langle 5^-|p_1+p_2|3^- \rangle \right. \nonumber \\
&&+2 (\br(1,5) \br(2,3) -\br(1,2) \br(3,5) )\sq(1,3) \sq(1,4) \sq(2,4) \sq(3,4) -
\br(2,5) \sq(1,4) \sq(2,4) \sq(3,4) (s_{15}+s_{25}) \nonumber \\
&&+2 \br(3,5) \sq(1,3) \sq(3,4) \langle 4^+|p_1p_5-p_2p_3|4^-\rangle
-2 \br(3,5) \sq(1,3) \sq(3,4) \langle 4^+|p_5(p_1+p_3)|4^-\rangle \nonumber \\
&&+\br(2,5) \sq(1,4) \sq(2,3) \langle 4^+|p_1p_5-p_2p_3|4^-\rangle
+\br(2,5) \sq(1,4) \sq(2,4) \langle 3^+|p_1p_5-p_2p_3|4^-\rangle \nonumber \\
&&\left. +\br(2,5) \br(3,5) \sq(3,4)  (\sq(1,5) \sq(2,4) \sq(3,4) +\sq(1,4) \sq(2,3) \sq(4,5) +
\sq(1,3) \sq(2,4) \sq(4,5) )+2 \br(1,5) ^2 \sq(1,3) \sq(1,4) ^2 \sq(4,5) \right] \nonumber \\
&&-\frac{\br(2,5) \sq(1,4) \sq(3,4) }{2 s_{12}s_{45}s_{3{\sss H}}} \langle 5^-|p_{\sss H}|3^-\rangle \nonumber \\
&&+\frac{1}{2 \br(3,5) s_{12} s_{45}s_{3{\sss H}}t_{125}} \left[
2 \br(1,2) \br(1,5) \br(3,5) \sq(1,3) ^2 \sq(1,4) \langle 5^-|p_1+p_2|4^- \rangle \right. \nonumber \\
&&-\br(1,2) \br(1,5) \br(2,5) \br(3,5) \sq(1,3) \sq(1,4) (\sq(1,4) \sq(2,3) +\sq(1,3) \sq(2,4) )+
\br(1,5) \br(3,5) \sq(1,3) \sq(1,4) \sq(3,4) \langle 2^-|2 p_5 p_1-3 p_1p_4|5^+\rangle \nonumber \\
&&+\br(1,5) \br(2,4) \sq(1,4) ^2 \sq(3,4) \langle 5^-|(p_1+2 p_2)p_3|5^+ \rangle -
\br(2,5) ^2 \br(3,5) \sq(1,4) \sq(2,3) \sq(3,4) (s_{12}+s_{14})  \nonumber \\
&&-2 \br(2,5) \sq(1,4) \sq(3,4)  \langle 5^-|p_1(p_5p_2p_3+p_3p_2p_4)|5^+ \rangle +
2\br(2,5) ^2 \sq(1,3) \sq(2,4) \langle 4^+|p_3(p_1p_4-p_5p_1)|5^+ \rangle \nonumber \\
&&+\br(2,4) \br(2,5) \sq(1,4) \sq(2,4) \sq(3,4)  \langle 5^-|(p_1- p_2)p_3|5^+ \rangle
-2 \br(2,5) ^3 \sq(2,4) \sq(3,4) \langle 1^+|p_5p_3-p_3p_4|2^- \rangle \nonumber \\
&&-2 \br(2,5) ^2 \sq(1,3) \sq(2,4) ^2 \langle 2^-|-p_1p_3+p_3p_4|5^+ \rangle
+\br(3,5) \br(4,5) \sq(1,4) \sq(3,4) ^2 \langle 2^-|p_1(p_2+p_4)-(p_4+p_5)p_1|5^+ \rangle \nonumber \\
&&+\br(2,5) \sq(1,4) \sq(3,4) \langle 5^-|p_4(p_1p_5-p_5p_2)p_3|5^+ \rangle
-\br(2,5) \br(3,5) \br(4,5) \sq(3,4) ^2 \sq(1,4) (s_{24}+s_{45})  \nonumber \\
&&\left. +2 \br(1,5) \br(2,5) \sq(1,4) \sq(3,4) \langle 1^+|p_2(p_4+p_5)p_3|5^+ \rangle +
2 \br(2,5) ^2 \sq(2,4) \sq(3,4) \langle 1^+|(p_5p_4+p_2p_5)p_3|5^+ \rangle \right] \nonumber \\
&&+\frac{\br(2,5) }{2 s_{25}s_{45}s_{3{\sss H}}t_{125}} \left[
2 \br(1,2) \sq(1,3) ^2 \sq(2,4) \langle 5^-|p_1+p_2|4^- \rangle \right. \nonumber \\
&&+2\sq(1,3) \sq(1,4) \sq(2,4) \sq(3,4) (\br(1,2) \br(4,5) -\br(1,5) \br(2,4) )+
\br(2,5) \sq(2,4) \sq(3,4) \langle 1^+|p_5(p_1+p_2)|3^- \rangle \nonumber \\
&&-2\sq(1,3) \sq(3,4) \langle 5^-|p_2(p_4p_2+p_5p_1)|4^- \rangle +
2 \br(4,5) \sq(1,4) \sq(3,4) \langle 4^+|p_2(p_4+p_5)|3^- \rangle  \nonumber \\
&&+\br(2,5) \sq(1,3) \sq(2,4) \langle 3^+|2p_5p_2+p_1p_5+p_2p_5|4^- \rangle
+\br(1,5) \sq(1,3) \sq(1,4) \langle 4^+|p_5(-2p_1-p_2)|3^- \rangle  \nonumber \\
&&\left. +\br(2,5) \sq(2,3) \sq(1,4) \langle 4^+|p_5(-2p_4-p_2)|3^- \rangle -
2 s_{45}\sq(1,3) \sq(3,4) \langle 5^-|2p_1+p_2|4^- \rangle  -
\br(4,5) \sq(1,4) \sq(3,4) ^2 (2s_{45} +s_{25}) \right] \nonumber \\
&&\left. +\frac{\br(2,5) ^2 \sq(1,3) \sq(2,4) }{s_{25}s_{45}s_{3{\sss H}}t_{245}}\left[
\langle 3^+|p_1(p_2+p_5)|4^- \rangle +t_{245} \sq(3,4) \right] \right\} \, ,
\label{sec:Aqqppm}
\end{eqnarray}
All the other sub-amplitudes can be obtained by relabelling and
by use of reflection symmetry, parity inversion and charge conjugation.

\subsection{Sub-amplitudes for Higgs $+$ one gluon and two $q\qb$ pairs}
\label{sec:appbc}

The colour decomposition of the amplitudes for Higgs, one gluon and two 
$q\bar q$ pairs is given in \eqn{FourQuarkGluonDecomp}, with $n$ = 5.
It can be explicitly re-written as,
\bea
\lefteqn{ {\cal M}_5(1_q, 2_{\bar{q}}; 3_Q, 4_{\Qb}; 5) }\nn\\
&=& g^3 {1\over \sqrt{2}} \Bigl[ \left( T^{a_5} \right)_{i_1}^{\;\ib_4} 
\delta_{i_3}^{\;\ib_2} m_{5,1}(1_q, 2_{\bar{q}}; 3_Q, 4_{\Qb}; 5) + 
\left( T^{a_5} \right)_{i_3}^{\;\ib_2} \delta_{i_1}^{\;\ib_4} 
m_{5,2}(1_q, 2_{\bar{q}}; 3_Q, 4_{\Qb}; 5) \label{fourquarkoneglue}\\
&& \qquad + {1\over N_c}
\left( T^{a_5} \right)_{i_1}^{\;\ib_2} \delta_{i_3}^{\;\ib_4} 
m_{5,3}(1_q, 2_{\bar{q}}; 3_Q, 4_{\Qb}; 5) + {1\over N_c}
\left( T^{a_5} \right)_{i_3}^{\;\ib_4} \delta_{i_1}^{\;\ib_2}
m_{5,4}(1_q, 2_{\bar{q}}; 3_Q, 4_{\Qb}; 5) \Bigr] \nn
\eea
For each sub-amplitude of type $m_{5,i}$, with $i=1, 2, 3, 4$, 
there are two independent
sub-amplitudes. The two independent sets of four sub-amplitudes
can be taken to be
\begin{eqnarray}
\lefteqn{m_{5,1}(1_q^+,2^-_{\bar{q}};3^+_Q,4^-_{\bar{Q}};5^+)=}\nonumber \\
&&iA \left\{ \frac{1}{\br(1,5) s_{12}s_{34}}\left[ \br(2,4) \sq(3,5)
(s_{12}-s_{15})-\br(2,4) \br(1,4) \sq(1,5) \sq(3,4) -\br(1,2) \sq(1,3) 
\langle 4^-|p_1+p_3|5^- \rangle \right] \right. \nonumber  \\
&&-\frac{1}{\br(1,5) s_{12}s_{34}t_{125}}\left[ -\br(1,2) ^2 \br(3,4)
\sq(1,3) ^2 \sq(2,5) -\br(1,4) \br(2,4) \sq(1,5) \sq(3,4)
(s_{15}+s_{25}) \right. \nonumber \\
&&\left. +s_{12}\br(2,4) ^2\sq(2,5) \sq(3,4) +\br(1,2) \br(3,4)
\sq(1,3) \sq(3,5) (s_{25}-s_{15})-s_{15} \br(2,5) \br(3,4) \sq(3,5) ^2 
\right] \nonumber \\
&&+\frac{1}{\br(1,5) s_{34}t_{125}} \left[ \br(2,4) \sq(4,3)
\langle 4^-|p_1+p_2|5^- \rangle +
\br(4,3) \sq(3,5) \langle 2^-|p_1+p_5|3^- \rangle \right] \nonumber   \\
&&-\frac{1}{\br(1,5) s_{12}s_{34}t_{345}} \left[
-\br(1,2) \sq(3,5) (s_{13}+s_{15}) \langle 4^-|p_3+p_5|1^- \rangle +
\br(2,4) ^2 \sq(1,2) \sq(3,4) \langle 1^-|p_3+p_4|5^- \rangle 
\right. \nonumber \\
&&\left. +\br(2,4) \sq(3,5) s_{12} (s_{35}+s_{45})+
\br(1,2) \br(1,4) \sq(1,3) \langle 1^+|(p_3+p_5)p_4|5^- \rangle \right]
\nonumber \\
&&-\frac{\br(1,4) }{\br(1,5) \br(4,5) s_{12} t_{345}} \left[
-\br(1,2) \sq(1,3) \langle 4^-|p_3+p_5|1^- \rangle +\br(2,4) \sq(1,2) 
\langle 2^-|p_4+p_5|3^- \rangle \right] \nonumber \\
&&-\frac{1}{s_{12}s_{34}s_{5{\sss H}}} \left[
-\br(1,2) \sq(1,5) ^2 \langle 4^-|p_1+p_2|3^- \rangle +
\sq(1,5) \sq(3,5) \langle 2^-|-p_1 p_3+p_3 p_1|4^+ \rangle
\right. \nonumber \\
&&\left. +\br(2,4) \sq(1,5) \sq(3,5) (s_{14}+s_{23}+s_{24})+
\br(3,4) \sq(3,5) ^2 \langle 2^-|p_3+p_4|1^- \rangle \right] \nonumber
\\
&&+\frac{\sq(1,3) }{s_{12}t_{123}s_{5{\sss H}}} \langle 2^-|p_1+p_3|5^-
\rangle \langle 4^-|p_{\sss H}|5^- \rangle \nonumber \\
&&\left. +\frac{\br(2,4) \sq(1,5) }{ s_{34}s_{5{\sss H}} }
\left[ \frac{\sq(5,1) \langle 1^-|p_2+p_4|3^- \rangle }{t_{234}}
+\sq(5,3) 
\right] \right\}
\label{sec:A1pmpmp}
\end{eqnarray}

\begin{eqnarray}
\lefteqn{m_{5,2}(1_q^+,2^-_{\bar{q}};3^+_Q,4^-_{\bar{Q}};5^+)=}\nonumber \\
&&iA \left\{-\frac{1}{\br(1,5) s_{12}s_{34}}\left[ \br(2,4) \sq(3,5)
(s_{12}-s_{15})-\br(2,4) \br(1,4) \sq(1,5) \sq(3,4) -\br(1,2) \sq(1,3) 
\langle 4^-|p_1+p_3|5^- \rangle \right] \right. \nonumber  \\
&&+\frac{1}{\br(1,5) s_{12}s_{34}t_{125}}\left[ -\br(1,2) ^2 \br(3,4)
\sq(1,3) ^2 \sq(2,5) -\br(1,4) \br(2,4) \sq(1,5) \sq(3,4)
(s_{15}+s_{25}) \right. \nonumber \\
&&\left. +s_{12}\br(2,4) ^2\sq(2,5) \sq(3,4) +\br(1,2) \br(3,4)
\sq(1,3) \sq(3,5) (s_{25}-s_{15})-s_{15} \br(2,5) \br(3,4) \sq(3,5) ^2 
\right] \nonumber \\
&&-\frac{\br(1,2) }{\br(1,5) \br(2,5) s_{34}t_{125}} \left[  \sq(1,3)
\langle 2^-|(p_1+p_5)p_3|4^+ \rangle -
\br(2,4)  \langle 3^+|p_4(p_2+p_5)|1^- \rangle \right] \nonumber   \\
&&+\frac{1}{\br(1,5) s_{12}s_{34}t_{345}} \left[
-\br(1,2) \sq(3,5) (s_{13}+s_{15}) \langle 4^-|p_3+p_5|1^- \rangle +
\br(2,4) ^2 \sq(1,2) \sq(3,4) \langle 1^-|p_3+p_4|5^- \rangle 
\right. \nonumber \\
&&\left. +\br(2,4) \sq(3,5) s_{12} (s_{35}+s_{45})+
\br(1,2) \br(1,4) \sq(1,3) \langle 1^+|(p_3+p_5)p_4|5^- \rangle \right]
\nonumber \\
&&-\frac{1}{\br(1,5) \br(3,5) s_{12}t_{345}}
\left[ (s_{13}+s_{15})  \langle 2^-|p_1(p_3+p_5)|4^+ \rangle
+\br(2,4) ^2 \sq(1,2)  \langle 1^-|(p_3+p_5)|4^- \rangle
\right. \nonumber \\
&& \left. +\br(2,4) s_{12}s_{35} \right] \nonumber \\
&&+\frac{1}{s_{12}s_{34}s_{5{\sss H}}} \left[
-\br(1,2) \sq(1,5) ^2 \langle 4^-|p_1+p_2|3^- \rangle +
\sq(1,5) \sq(3,5) \langle 2^-|-p_1 p_3+p_3 p_1|4^+ \rangle
\right. \nonumber \\
&&\left. +\br(2,4) \sq(1,5) \sq(3,5) (s_{14}+s_{23}+s_{24})+
\br(3,4) \sq(3,5) ^2 \langle 2^-|p_3+p_4|1^- \rangle \right] \nonumber
\\
&&+\frac{\br(2,4) \sq(3,5) }{s_{12}t_{124}s_{5{\sss H}}}\left[
\sq(1,5) t_{124}+\sq(3,5) \langle 3^-|p_2+p_4|1^- \rangle \right]
\nonumber \\
&&\left. -\frac{\sq(1,3) }{s_{34}t_{134}s_{5{\sss H}}} \langle 4^-|p_1+p_3|5^-
\rangle \langle 2^-|p_{\sss H}|5^- \rangle \right\} \, ,
\label{sec:A2pmpmp}
\end{eqnarray}

\begin{eqnarray}
\lefteqn{m_{5,3}(1_q^+,2^-_{\bar{q}};3^+_Q,4^-_{\bar{Q}};5^+)=}\nonumber \\
&&iA \left\{ -\frac{1}{\br(1,5) s_{34}t_{125}} \left[ \br(2,4) \sq(4,3)
\langle 4^-|p_1+p_2|5^- \rangle +
\br(4,3) \sq(3,5) \langle 2^-|p_1+p_5|3^- \rangle \right] \right. \nonumber   \\
&&+\frac{\br(1,2) }{\br(1,5) \br(2,5) s_{34}t_{125}} \left[  \sq(1,3)
\langle 2^-|(p_1+p_5)p_3|4^+ \rangle -
\br(2,4)  \langle 3^+|p_4(p_2+p_5)|1^- \rangle \right] \nonumber   \\
&&+\frac{\sq(1,3) }{s_{34}t_{134}s_{5{\sss H}}} \langle 4^-|p_1+p_3|5^-
\rangle \langle 2^-|p_{\sss H}|5^- \rangle \nonumber \\
&&\left. -\frac{\br(2,4) \sq(1,5) }{ s_{34}s_{5{\sss H}} }
\left[ \frac{\sq(5,1) \langle 1^-|p_2+p_4|3^- \rangle }{t_{234}}+\sq(5,3)
\right] \right\} \, ,
\label{sec:A3pmpmp}
\end{eqnarray}

\begin{eqnarray}
\lefteqn{m_{5,4}(1_q^+,2^-_{\bar{q}};3^+_Q,4^-_{\bar{Q}};5^+)=}\nonumber \\
&&iA \left\{ \frac{1}{\br(1,5) \br(3,5) s_{12}t_{345}}
\left[ (s_{13}+s_{15})  \langle 2^-|p_1(p_3+p_5)|4^+ \rangle
+\br(2,4) ^2 \sq(1,2)  \langle 1^-|(p_3+p_5)|4^- \rangle
\right. \right.\nonumber \\
&& \left. +\br(2,4) s_{12}s_{35} \right] \nonumber \\
&&+\frac{\br(1,4) }{\br(1,5) \br(4,5) s_{12} t_{345}} \left[
-\br(1,2) \sq(1,3) \langle 4^-|p_3+p_5|1^- \rangle +\br(2,4) \sq(1,2) 
\langle 2^-|p_4+p_5|3^- \rangle \right] \nonumber \\
&&-\frac{\sq(1,3) }{s_{12}t_{123}s_{5{\sss H}}} \langle 2^-|p_1+p_3|5^-
\rangle \langle 4^-|p_{\sss H}|5^- \rangle \nonumber \\
&&\left. -\frac{\br(2,4) \sq(3,5) }{s_{12}t_{124}s_{5{\sss H}}}\left[
\sq(1,5) t_{124}+\sq(3,5) \langle 3^-|p_2+p_4|1^- \rangle \right] \right\} \, .
\label{sec:A4pmpmp}
\end{eqnarray}

\begin{eqnarray}
\lefteqn{m_{5,1}(1_q^+,2^-_{\bar{q}};3^-_Q,4^+_{\bar{Q}};5^+)=}\nonumber\\
&&iA \left\{ \frac{1}{\br(1,5) s_{12}s_{34}t_{345}}\left[
\br(2,3) \sq(1,2) (- \br(2,3) \sq(4,5) \langle 1^-|p_4+p_5|3^- \rangle
+\br(1,3) \sq(3,5) \langle 2^-|p_3+p_5|4^- \rangle )\right. \right. \nonumber \\
&& \left. +\br(1,2) \langle 3^-|p_4+p_5|1^- \rangle (- \br(1,3) \sq(1,4) 
\sq(3,5) +\sq(4,5) (s_{14}+s_{15}))-s_{12}s_{45}\br(2,3) \sq(4,5)
\right] \nonumber \\
&&-\frac{1}{s_{12}s_{34}s_{5{\sss H}}} \left[ 
-\br(1,2) \sq(1,5) ^2 \langle 3^-|p_1+p_2|4^- \rangle
\right. \nonumber \\
&&\left. -\sq(1,5) \sq(4,5) (2 \langle 2^-|p_1  p_4|3^+ \rangle -
\br(2,3) (s_{13}+s_{14}+s_{23}+s_{24}))-\br(3,4) \sq(4,5) ^2 \langle 2^-|p_3+p_4|1^- \rangle
\right] \nonumber \\
&&+\frac{1}{\br(1,5) s_{12} s_{34}t_{125}} \left[
-\br(1,2) ^2\br(3,4) \sq(1,4) ^2 \sq(2,5) +
\br(1,2) (-\br(2,3) ^2 \sq(1,2) \sq(2,5) \sq(3,4) +
\br(3,4) \sq(1,4) \sq(4,5) (s_{25}-s_{15})) \right. \nonumber \\
&&\left. +\sq(1,5) (-\br(1,3) \br(2,3) \sq(3,4) (s_{15}+s_{25})+
\br(1,5) \br(2,5) \br(3,4) \sq(4,5) ^2) \right] \nonumber \\
&&+\frac{1}{\br(1,5) s_{12}s_{34}} \left[ \br(2,3) \sq(1,5) 
\langle 1^-|p_3-p_5|4^- \rangle +\br(1,2) (\br(1,3) \sq(1,4) \sq(1,5) -
\sq(4,5) \langle 3^-|p_2-p_4|1^- \rangle ) \right] \nonumber \\
&&-\frac{1}{\br(1,5) \br(4,5) s_{12}t_{345}} \left[
-\br(2,3) ^2 \sq(1,2) \langle 1^-|p_4+p_5|3^- \rangle +
\br(1,2) (s_{14}+s_{15}) \langle 3^-|p_4+p_5|1^- \rangle -
s_{12}s_{45}\br(2,3) \right] \nonumber \\
&&+\frac{\br(2,3) \sq(4,5) }{s_{12}t_{123}s_{5{\sss H}}} \left[
-\sq(1,5) t_{123}-\sq(4,5) \langle 4^-|p_2+p_3|1^- \rangle \right ]
\nonumber \\
&&+\frac{\br(2,3) \sq(1,5) }{s_{34} t_{234}s_{5{\sss H}}}\left[
-\sq(1,5) \langle 1^-|p_2+p_3|4^- \rangle -\sq(4,5) t_{234} \right]
\nonumber \\
&&\left. +\frac{1}{\br(1,5) s_{34} t_{125}} \left[ \br(2,3) \sq(3,4) 
\langle 3^-|p_1+p_2|5^- \rangle
+\br(3,4) \sq(4,5) \langle 2^-|p_1+p_5|4^- \rangle \right] \right\} \, ,
\label{sec:A1pmmpp}
\end{eqnarray}

\begin{eqnarray}
\lefteqn{m_{5,2}(1_q^+,2^-_{\bar{q}};3^-_Q,4^+_{\bar{Q}};5^+)=}\nonumber\\
&&iA \left\{ -\frac{1}{\br(1,5) s_{12}s_{34}t_{345}}\left[
\br(2,3) \sq(1,2) (- \br(2,3) \sq(4,5) \langle 1^-|p_4+p_5|3^- \rangle
+\br(1,3) \sq(3,5) \langle 2^-|p_3+p_5|4^- \rangle )\right. \right. \nonumber \\
&& \left. +\br(1,2) \langle 3^-|p_4+p_5|1^- \rangle (- \br(1,3) \sq(1,4) 
\sq(3,5) +\sq(4,5) (s_{14}+s_{15}))-s_{12}s_{45}\br(2,3) \sq(4,5)
\right] \nonumber \\
&&+\frac{1}{s_{12}s_{34}s_{5{\sss H}}} \left[ 
-\br(1,2) \sq(1,5) ^2 \langle 3^-|p_1+p_2|4^- \rangle
\right. \nonumber \\
&&\left. -\sq(1,5) \sq(4,5) (2 \langle 2^-|p_1  p_4|3^+ \rangle -
\br(2,3) (s_{13}+s_{14}+s_{23}+s_{24}))-\br(3,4) \sq(4,5) ^2 \langle 2^-|p_3+p_4|1^- \rangle
\right] \nonumber \\
&&-\frac{1}{\br(1,5) s_{12} s_{34}t_{125}} \left[
-\br(1,2) ^2\br(3,4) \sq(1,4) ^2 \sq(2,5) +
\br(1,2) (-\br(2,3) ^2 \sq(1,2) \sq(2,5) \sq(3,4) +
\br(3,4) \sq(1,4) \sq(4,5) (s_{25}-s_{15})) \right. \nonumber \\
&&\left. +\sq(1,5) (-\br(1,3) \br(2,3) \sq(3,4) (s_{15}+s_{25})+
\br(1,5) \br(2,5) \br(3,4) \sq(4,5) ^2) \right] \nonumber \\
&&-\frac{1}{\br(1,5) s_{12}s_{34}} \left[ \br(2,3) \sq(1,5) 
\langle 1^-|p_3-p_5|4^- \rangle +\br(1,2) (\br(1,3) \sq(1,4) \sq(1,5) -
\sq(4,5) \langle 3^-|p_2-p_4|1^- \rangle ) \right] \nonumber \\
&&-\frac{\sq(1,4) }{s_{12}t_{124}s_{5{\sss H}}} \langle 2^-|p_1+p_4|5^-
\rangle \langle 3^-|p_{\sss H}|5^- \rangle \nonumber \\
&&+\frac{\br(1,3) }{\br(1,5) \br(3,5) s_{12}t_{345}}\left[
-\br(1,2) \sq(1,4) \langle 3^-|p_4+p_5|1^- \rangle +
\br(2,3) \sq(1,2) \langle 2^-|p_3+p_5|4^- \rangle \right] \nonumber \\
&&-\frac{\br(1,2) }{\br(1,5) \br(2,5) s_{34}t_{125}} \left[ \br(2,3)
\sq(3,4) \langle 3^-|p_2+p_5|1^- \rangle -
\br(3,4) \sq(1,4) \langle 2^-|p_1+p_5|4^- \rangle \right] \nonumber \\
&&\left. -\frac{\sq(1,4) }{s_{34}t_{134}s_{5{\sss H}}} \langle 2^-|p_{\sss H}|5^- \rangle
\langle 3^-|p_1+p_4|5^- \rangle \right\} \, ,
\label{sec:A2pmmpp}
\end{eqnarray}

\begin{eqnarray}
\lefteqn{m_{5,3}(1_q^+,2^-_{\bar{q}};3^-_Q,4^+_{\bar{Q}};5^+)=}\nonumber\\
&&iA \left\{ -\frac{\br(2,3) \sq(1,5) }{s_{34} t_{234}s_{5{\sss H}}}\left[
-\sq(1,5) \langle 1^-|p_2+p_3|4^- \rangle -\sq(4,5) t_{234} \right] \right. 
\nonumber \\
&&-\frac{1}{\br(1,5) s_{34} t_{125}} \left[ \br(2,3) \sq(3,4) 
\langle 3^-|p_1+p_2|5^- \rangle
+\br(3,4) \sq(4,5) \langle 2^-|p_1+p_5|4^- \rangle \right] \nonumber  \\
&&+\frac{\br(1,2) }{\br(1,5) \br(2,5) s_{34}t_{125}} \left[ \br(2,3)
\sq(3,4) \langle 3^-|p_2+p_5|1^- \rangle -
\br(3,4) \sq(1,4) \langle 2^-|p_1+p_5|4^- \rangle \right] \nonumber \\
&&\left. +\frac{\sq(1,4) }{s_{34}t_{134}s_{5{\sss H}}} \langle 2^-|p_{\sss H}|5^- \rangle
\langle 3^-|p_1+p_4|5^- \rangle \right\} \, ,
\label{sec:A3pmmpp}
\end{eqnarray}

\begin{eqnarray}
\lefteqn{m_{5,4}(1_q^+,2^-_{\bar{q}};3^-_Q,4^+_{\bar{Q}};5^+)=}\nonumber\\
&&iA \left\{ \frac{\sq(1,4) }{s_{12}t_{124}s_{5{\sss H}}} \langle 2^-|p_1+p_4|5^-
\rangle \langle 3^-|p_{\sss H}|5^- \rangle \right. \nonumber \\
&&-\frac{\br(1,3) }{\br(1,5) \br(3,5) s_{12}t_{345}}\left[
-\br(1,2) \sq(1,4) \langle 3^-|p_4+p_5|1^- \rangle +
\br(2,3) \sq(1,2) \langle 2^-|p_3+p_5|4^- \rangle \right] \nonumber \\
&&+\frac{1}{\br(1,5) \br(4,5) s_{12}t_{345}} \left[
-\br(2,3) ^2 \sq(1,2) \langle 1^-|p_4+p_5|3^- \rangle +
\br(1,2) (s_{14}+s_{15}) \langle 3^-|p_4+p_5|1^- \rangle
\right. \nonumber \\ 
&& \left.  -
s_{12}s_{45}\br(2,3) \right] \nonumber \\
&&\left. -\frac{\br(2,3) \sq(4,5) }{s_{12}t_{123}s_{5{\sss H}}} \left[
-\sq(1,5) t_{123}-\sq(4,5) \langle 4^-|p_2+p_3|1^- \rangle \right] \right\}\, .
\label{sec:A4pmmpp}
\end{eqnarray}
All the other sub-amplitudes can be obtained by relabelling and
by use of parity inversion, reflection symmetry and charge 
conjugation.
For identical quarks, we must subtract
from \eqn{FourQuarkGluonDecomp} the same term with the antiquarks (but not the
quarks) exchanged $(2_\qb\leftrightarrow 4_{\overline{Q}})$.

\subsection{Colour matrices for Higgs $+$ five partons}
\label{sec:appc}

Using \eqn{GluonDecompNew}, the squared amplitude for Higgs $+\ n$ 
gluons, summed over colours, is
\beq
\sum_{\rm colours} |{\cal M}_n(1,\dots,n)|^2 
= 2^{n-4} (g^2)^{n-2} \sum_{i,j=1}^{(n-2)!} 
\tilde{c}_{ij}\ m_i\ (m_j)^* \, , \label{square}
\eeq
where the subscript $i$ on $m_i$ now refers to the sub-amplitude 
$m_n(1,\ldots,n)$ evaluated for the  $i^{\rm th}$ permutation $P_i$ 
in $S_{n-2}$. The colour matrix $\tilde{c}_{ij}$ is,
\begin{equation}
\tilde{c}_{ij} = \sum_{\rm colours} \left(P_i \{ F^{a_2} \cdots 
F^{a_{n-1}}\} \right)_{a_1 a_n}
\left[\left(P_j \{F^{a_2} \cdots F^{a_{n-1}}\}\right)_{a_1 a_n}
 \right]^*\, ,\label{ctilde}
\end{equation}
where we have re-written the structure constants in terms of the
$SU(N_c)$ generators in the adjoint representation,
$(F^a)_{bc}\equiv i f^{bac}$. For $n=5$ the colour matrix is,
\beq
\tilde{c}_{ij}= \frac{\CF N_c^4 }{2} 
\left(
\begin{array}{cccccccccccc}
4 && 2 && 2 && 1 && 1 && 0 \\
2 && 4 && 1 && 0 && 2 && 1 \\
2 && 1 && 4 && 2 && 0 && 1 \\
1 && 0 && 2 && 4 && 1 && 2 \\
1 && 2 && 0 && 1 && 4 && 2 \\
0 && 1 && 1 && 2 && 2 && 4
\end{array}
\right)
, \hspace{1cm} 
m_i=\left(
\begin{array}{c}
(1,2,3,4,5)\\
(1,2,4,3,5)\\
(1,3,2,4,5)\\
(1,3,4,2,5)\\
(1,4,2,3,5)\\
(1,4,3,2,5)
\end{array}
\right) .
\eeq

Using \eqn{TwoQuarkGluonDecomp}, the squared amplitude for Higgs $+\ (n-2)$
gluons and a $q\qb$ pair, summed over colours, is
\beq
\sum_{\rm colours} |{\cal M}_n(q,\bar{q}; 3,\ldots,n)|^2
  = 2^{n-4} (g^2)^{n-2} \sum_{i,j=1}^{(n-2)!} 
d_{ij} m_i (m_j)^* \, , \label{square2}
\end{equation}
where, as in \eqn{square}, the subscript $i$ on $m_i$ now refers to the 
sub-amplitude $m_n(1_q,2_{\bar{q}}; 3,\ldots,n)$ evaluated for the
$i^{\rm th}$ permutation $P_i$ in $S_{n-2}$.
The colour matrix $d_{ij}$ is,
\beq
d_{ij} = \sum_{\rm colours} 
\left(P_i \{T^{a_3}\ldots T^{a_n}\} \right)_{i}^{~\jb}\ \left[ 
\left(P_j \{T^{a_3}\ldots T^{a_n}\} \right)_{i}^{~\jb} \right]^\dagger\,
.\label{dcolour}
\eeq
For $n=5$ it is,
\beq
d_{ij} = \CF  
\left(
\begin{array}{cccccccccccc}
a && b && b && c && c && d \\
b && a && c && d && b && c \\
b && c && a && b && d && c \\
c && d && b && a && c && b \\
c && b && d && c && a && b \\
d && c && c && b && b && a 
\end{array}
\right)
, \hspace{1cm} 
m_i=\left(
\begin{array}{c}
(1,2;3,4,5)\\
(1,2;3,5,4)\\
(1,2;4,3,5)\\
(1,2;4,5,3)\\
(1,2;5,3,4)\\
(1,2;5,4,3)
\end{array}
\right) ,
\eeq
where $a=N_c \CF^2$, $b=-\CF/2$, $c=1/(4N_c)$ and $d=(N_c^2+1)/(4N_c)$.

Using \eqn{fourquarkoneglue}, the squared amplitude for Higgs $+$ one
gluon and two $q\qb$ pairs, summed over colours, is
\bea
\lefteqn{
\sum_{\rm colours} |{\cal M}_5(1_q, 2_{\bar{q}}; 3_Q, 4_{\Qb}; 5)|^2 }
\label{square3}\\
&=& {g^6\over 2}\ \CF \left[ N_c^2 \left( |m_{5,1}|^2 + |m_{5,2}|^2 \right)
+ |m_{5,3}|^2 + |m_{5,4}|^2 + 2\ {\rm Re} \left( m_{5,1}^* + m_{5,2}^* \right)
\left( m_{5,3} + m_{5,4} \right) \right]\, ,\nn
\eea
with $m_{5,i} = m_{5,i}(1_q, 2_{\bar{q}}; 3_Q, 4_{\Qb}; 5)$ and
$i=1,\ldots,4$. For identical quarks, we use \eqn{fourquarkoneglue} and
subtract the same term with the antiquarks (but not the
quarks) exchanged $(2_\qb\leftrightarrow 4_{\overline{Q}})$.
Using the short-hand notations, 
$m_{5,i}(2,4) = m_{5,i}(1_q, 2_{\bar{q}}; 3_q, 4_{\qb}; 5)$ and
$m_{5,i}(4,2) = m_{5,i}(1_q, 4_{\bar{q}}; 3_q, 2_{\qb}; 5)$, the
squared amplitude summed over colours is
\bea
\lefteqn{
\sum_{\rm colours} |{\cal M}_5(1_q, 2_{\bar{q}}; 3_q, 4_{\qb}; 5)|^2 }
\nn\\ &=& {g^6\over 2}\ \CF \biggl\{
N_c^2 \sum_{i=1}^2 \left( |m_{5,i}(2,4)|^2 + |m_{5,i}(4,2)|^2 \right) 
+ \sum_{i=3}^4 \left( |m_{5,i}(2,4)|^2 + |m_{5,i}(4,2)|^2 \right) \nn\\
&& \qquad \quad - 2 N_c\ {\rm Re}\ \left[ 
m_{5,1}^*(2,4) (m_{5,1}(4,2) + m_{5,2}(4,2) + m_{5,3}(4,2)) +
m_{5,1}^*(4,2) m_{5,3}(2,4) \right. \nn\\
&& \qquad \qquad \qquad \quad \left. +
m_{5,2}^*(2,4) (m_{5,1}(4,2) + m_{5,2}(4,2) + m_{5,4}(4,2) ) +
m_{5,2}^*(4,2) m_{5,4}(2,4) \right]  \nn\\
&& \qquad \quad + 2\ {\rm Re}\ \left[ 
(m_{5,1}^*(2,4) + m_{5,2}^*(2,4)) (m_{5,3}(2,4) + m_{5,4}(2,4))
\right. \nn\\ && \qquad \qquad \qquad \left. +
(m_{5,1}^*(4,2) + m_{5,2}^*(4,2)) (m_{5,3}(4,2) + m_{5,4}(4,2)) \right]
\nn\\ && \qquad \quad  
- {2\over N_c} {\rm Re}\  \left[ (m_{5,3}^*(2,4) + m_{5,4}^*(2,4))
(m_{5,3}(4,2) + m_{5,4}(4,2)) \right]  \biggr\} \, .\label{idsquare}
\eea

\end{document}